\documentclass[letterpaper,11pt,oneside]{article}

\usepackage[T1]{fontenc}
\usepackage[utf8]{inputenc}
\usepackage[margin=1in]{geometry}
\usepackage{microtype}
\usepackage{amsmath,amssymb,amsthm,bm}
\usepackage{graphicx}
\usepackage{booktabs}
\usepackage{tabularx}
\usepackage{makecell}
\usepackage{array}
\usepackage{enumitem}
\usepackage{indentfirst}
\usepackage[authoryear,round]{natbib}
\usepackage{authblk}
\usepackage[hidelinks]{hyperref}
\hypersetup{
  pdftitle={Proxy-Adjusted Causal Discovery from Targeted Interventions},
  pdfauthor={Li Chen, Wei Pan, and Xiaotong Shen}
}
\bibpunct{(}{)}{;}{a}{,}{,}
\setlist[itemize]{leftmargin=1.5em}
\setlist[enumerate]{leftmargin=1.5em}

\newcommand{\E}{\mathbb{E}}
\newcommand{\Pbb}{\mathbb{P}}

\newcommand{\Pa}[1]{{\operatorname{pa}(#1)}}
\newcommand{\CLLR}{{conditional log-likelihood ratio}}

\theoremstyle{plain}
\newtheorem{theorem}{Theorem}

\newtheorem{proposition}[theorem]{Proposition}
\newtheorem{remark}[theorem]{Remark}
\newtheorem{corollary}[theorem]{Corollary}
\newtheorem{definition}[theorem]{Definition}
\theoremstyle{definition}
\newtheorem{assumption}[theorem]{Assumption}

\title{Proxy-Adjusted Causal Discovery from Targeted Interventions}
\author[1]{Li Chen}
\author[2]{Wei Pan}
\author[1]{Xiaotong Shen\thanks{Corresponding author: \texttt{xshen@umn.edu}.}}
\affil[1]{School of Statistics, University of Minnesota\\Minneapolis, MN 55455, USA}
\affil[2]{Division of Biostatistics and Health Data Science\\School of Public Health, University of Minnesota\\Minneapolis, MN 55455, USA}

\date{}

\begin{document}
\maketitle

\begin{abstract}
Randomized perturbations can reveal downstream responses without identifying
which causal relationships are direct. Conditioning on intermediate responses
may induce associations through unmeasured common causes, even when the source
is randomized. We introduce Proxy-Adjusted Balanced Masking (PABM), a
nonparametric framework for recovering directed acyclic graphs from targeted
interventions, response data, and recorded proxies. PABM uses valid
target-specific interventions to identify candidate ancestors. It then
compares conditional target distributions with and without a candidate
response and, when supported by the design, its intervention variable,
adjusting for other ancestors and proxies. Each comparison uses matched
observations, adjustment variables, and fitting procedures. Identification
requires every included nonparent comparison to have zero conditional gain and
each parent to have positive gain in at least one comparison. We establish
population identification, finite-sample recovery conditions that accommodate
incomplete adjustment and estimation error, and familywise error control under
valid held-out p-values. Continuous-response simulations show favorable graph
recovery relative to specified comparator pipelines; K562-calibrated count
simulations reveal a selection--ranking tradeoff. Proxy ablations assess
sensitivity to recorded information. An analysis of K562 Perturb-seq
data illustrates descriptive candidate-network construction; proxy adequacy
for unmeasured biological variation remains unresolved.
\end{abstract}

\par\medskip

\section{Introduction}
{Perturbation experiments measure how a system responds when selected components are
changed. Our goal is to determine which components directly affect others, rather
than merely transmit an effect through intermediate variables. Perturb-seq provides
a motivating example: CRISPR interference (CRISPRi) reduces expression of selected genes, and
single-cell ribonucleic acid (RNA) sequencing measures the resulting expression profiles
\citep{replogle2022}. A response to an intervention on one gene can arise either through
a direct effect on another gene or through a chain of intermediate genes. We seek
a directed acyclic graph (DAG), a graph without directed cycles, whose edges represent direct causal relationships among
the modeled responses \citep{pearl2009causality,spirtes2001cps}.
A direct cause is called a \emph{parent}; a variable with a directed path to the
target is an \emph{ancestor}. A strict ancestor excludes the
target itself, and a nonparent lacks a direct edge to the target.}

{The difficulty persists even when the source variable is randomized directly.
Suppose \(X_j\) affects \(X_r\) only through an intermediate response \(M\), and an
unmeasured cell-state variable \(H\) affects both \(M\) and \(X_r\):
\[
X_j\longrightarrow M\longrightarrow X_r,\qquad
H\longrightarrow M,\qquad H\longrightarrow X_r.
\]
Conditioning on \(M\) blocks the indirect path, but it can also make \(X_j\)
associated with \(H\): among cells with the same \(M\), a larger contribution from
\(X_j\) can be offset by a smaller contribution from \(H\). Because \(H\) also
affects \(X_r\), the adjusted comparison can suggest a direct relationship where
none exists. This is conditioning on a common effect, often called collider bias.
It does not require imperfect knockdown. In the assignment model
\(U_j\to X_j\), the same conditioning can also associate the randomized assignment
\(U_j\) with \(H\).
Conditioning on an observed response is distinct from experimentally setting
its value.}

In an additive linear structural model, valid instrumental variables
(IVs), such as suitable randomized interventions, can identify direct-effect
coefficients using regressors predicted from the instruments. This requires
enough independent instrumental variation to distinguish the coefficients of
\(X_j\) and \(M\). Simply substituting predicted means is insufficient for
general nonlinear relationships: a quadratic term in \(M\), for example,
requires \(E(M^2\mid Z)\), rather than \(\{E(M\mid Z)\}^2\), where
\(Z\) denotes the valid instruments. We seek graph recovery without
specifying these structural functions.

{We study whether recorded auxiliary measurements that carry information about
the unmeasured causes, called proxies, can support this adjustment. Our method,
Proxy-Adjusted Balanced Masking (PABM), provides a nonparametric framework:
the structural functions are not restricted to a prespecified parametric
family. PABM first uses target-specific interventions to identify candidate
upstream variables. This step requires intervention variables independent of relevant unmeasured causes
after adjustment, effects confined to the assigned target's mechanism,
remaining variation in those intervention variables, and detectable propagation along the
relevant paths. PABM then adjusts for other upstream responses and the recorded
measurements, comparing the target distribution with and without the candidate
response and, when supported by the design, its recorded intervention variable. \emph{Masking} means
omitting the candidate variable; \emph{balanced} means that its addition is the only
change in the data inputs and fitting protocol within each comparison.
The reported implementations use Gaussian or Poisson working
distributions with estimated conditional means. Detecting changes in variance
or distributional shape requires richer fitted distributions and a suitable
ancestor screen, as discussed in Section~\ref{subsec:pairwise-discovery}.}

{For these comparisons to identify direct edges, every comparison used for an
indirect candidate must show conditional independence from the target after adjustment,
while each direct parent must remain detectable in at least one comparison.
A measurement's correlation with cell state does not by itself ensure this
requirement. We state the identifying conditions and quantify sensitivity to
incomplete adjustment. In the K562 illustration, the recorded block contains
experimental-group indicators and mitochondrial-read percentage; whether these
adequately account for the relevant unmeasured biological variation is unresolved.
The analysis therefore illustrates construction and descriptive assessment of a
candidate network.}

Standard observational procedures provide important baselines, while this
task requires additional intervention and proxy conditions. {The PC algorithm, which uses conditional-independence tests, Greedy Equivalence}
Search, NOTEARS, Structural Agnostic Modeling, and Masked Gradient-Based Causal
Structure Learning
are commonly formulated {under causal sufficiency, the absence of unmeasured common causes among
the modeled variables}
\citep{kalisch2007pc,chickering2002ges,zheng2018notears,kalainathan2022sam,ng2022mcsl}.
Fast Causal Inference allows latent variables {but generally represents multiple causal graphs compatible with the
observed conditional independences, leaving some directions unresolved} \citep{zhang2008fci}. Intervention-based structural
learning and invariant prediction add directional information
\citep{peters2016invariant,brouillard2020dcdi,lippe2022enco}, yet exact recovery
under unmeasured confounding still requires assumptions connecting the
assignment mechanism, latent variables, and direct structural relations.
Nonlinear hidden-confounder procedures instead impose structured equations or
latent-error models, such as correlated Gaussian errors in DeFuSE and a
nonlinear latent-confounding {model} in NOCADILAC
\citep{li2024nonlinear,kaltenpoth2023nocadilac}.

{IV methods use intervention variables that change a source
variable, are independent of relevant unmeasured causes after adjustment,
and affect other responses only through that source. These requirements are
called relevance, conditional exogeneity, and exclusion, respectively.} GrIVET
combines uncertain additive interventions with a Gaussian DAG,
screening ancestral relations before estimating direct effects with candidate
instruments; GAMPI {uses a related strategy for identifying ancestors before direct effects
with generalized-linear}
structural equations \citep{chen2024grivet,wang2024gampi}. For Perturb-seq,
ARGEN \citep{park2026argen} uses differential-expression information to recover
descendants and a two-stage quasi-likelihood procedure to recover parents. {It constructs a model-based expression predictor in a first stage from
intervention variables and observed covariates, with the intervention variables serving as instruments.
This construction uses a count measurement model and a log-linear
structural model.} These procedures
{identify direct edges through structural coefficients under their
specified models.}

Nonparametric IV methods also estimate unknown structural functions
by requiring, for example, that the structural error have mean zero
conditional on the instruments, together with restrictions ensuring
identification \citep{newey2003npiv}. PABM targets graph structure through
proxy-adjusted conditional-distribution comparisons under the intervention,
conditional-independence, and detectability requirements described above.

{A second line of work uses recorded proxies and restrictions on
their conditional distributions. {Nonparametric proxy-based tests address a local causal null or the causal
direction between two variables through restrictions linking the distributions
of recorded proxies, hidden causes, and outcomes}
\citep{liu2024proxyci,wu2025bivariateproxy}. {Proximal methods identify causal effects using additional restrictions
on these distributions} \citep{miao2018proxy,cui2024proximal}, and
linear proxy-selection methods address causal-effect estimation
\citep{xie2024autoproxy}. The confounder-blanket framework recovers causal
ordering from background covariates under restrictions on their graphical
relation to the responses \citep{watson2022confounderblanket}. These approaches
establish that flexible inference with proxy information is possible under
appropriate identifying conditions. Our question is how recorded proxies and
target-specific interventions can jointly support direct-parent discovery
within an ancestor family.}

Recent perturbation-specific procedures estimate linear DAGs {from hard
interventions, which replace a source variable's generating mechanism,
for example by setting its value externally. Other procedures infer cyclic networks,} invert intervention-response matrices,
model latent Gaussian DAGs with Poisson measurements, or use Bayesian and
likelihood-based interventional graph models
\citep{xue2025dotears,rohbeck2024bicycle,brown2025inspre,zhang2026latentcount,
han2025ibcd,vinas2026pacer}.
{The diversity of these targets and models motivates a formulation
that separates the identifying restrictions from the choice of regression
model used to estimate each conditional distribution.}

{Conditional predictive impact provides a related predictive-relevance
construction \citep{watson2021cpi}; PABM's causal interpretation additionally
relies on the ancestor restriction and comparison-specific separation
conditions.}

The contributions are threefold.
\begin{enumerate}[label=(\roman*),leftmargin=1.8em]
\item PABM combines target-specific interventions and recorded proxies in a
direct-parent identification framework. For an eligible nonparent, every
available comparison must satisfy its conditional-independence null; a parent
may be detected by either the response or intervention-assignment comparison.
This joint-null/either-alternative structure accommodates general target
distributions under strong, pair-specific proxy-adjustment requirements.
\item The analysis separates residual dependence after proxy adjustment from
working-family approximation, conditional-distribution estimation, and
held-out validation error. It gives population identification, finite-sample
graph-recovery conditions, and familywise error rate (FWER) control when the comparisons
selected in training have valid held-out p-values. Sample splitting and
multiple-testing adjustment alone do not establish that validity.
\item Gaussian, continuous-response, and K562-calibrated count simulations
examine selection, calibration, graph recovery, and sensitivity to recorded
proxy information. A K562 Perturb-seq analysis illustrates descriptive
candidate-network construction without a reference DAG. The experiments
distinguish the proved specializations from the empirical adaptive
implementation and reveal a selection--ranking tradeoff in the count setting.
\end{enumerate}
The pair-specific requirements neither imply nor follow generally from the
structural and measurement assumptions of model-based IV methods. They also
distinguish a recorded proxy for unmeasured confounding from ARGEN's
model-generated expression predictor. The framework applies more broadly to
targeted-intervention studies with linked pre-intervention covariates or
auxiliary measurements whose causal role supports the required adjustment.

Section~2 defines the statistical model, graph targets, and population
identification conditions. Section~3 develops the PABM procedure. Section~4
gives graph-recovery and calibration guarantees. Section~5 presents the
numerical studies and Perturb-seq analysis, and Section~6 discusses the scope
of the resulting framework.

\section{Data, Identifying Conditions, and Causal Targets}

The data provide two complementary forms of information. Valid interventions can
identify which responses are upstream of a target; recorded proxies can help
separate direct parents from indirect ancestors. This section states the
conditions required for these two roles and defines the conditional-likelihood
gains that PABM estimates. The structural DAG is the primary target. Assignment
comparisons also describe which of its edges respond under the observed design.

\subsection{Structural Model and Graph-Recovery Target}

Consider observed responses, intervention assignments, recorded
covariates and proxy measurements, and unmeasured common causes. The following
notation separates these four roles and defines the graph-recovery target.
Let \([p]=\{1,\ldots,p\}\), let
\(X=(X_1,\ldots,X_p)\) denote the variables whose causal relations are of
interest, and let \(H=(H_1,\ldots,H_q)\) collect shared unmeasured
confounders that may affect several components of \(X\). Let \(W\)
denote the recorded covariate and proxy block. For each \(r\in[p]\), let \(W_r\)
denote the possibly empty subvector whose observed inputs enter the mechanism
for \(X_r\), and suppose
\begin{equation}
X_r = f_r\bigl(X_{\Pa{r}}, H, U_r, W_r, \varepsilon_r\bigr),
\label{eq:structural-system}
\end{equation}
where \(\Pa{r}\subseteq[p]\setminus\{r\}\) is the minimal observed
parent set of \(X_r\), \(U_r\) is its observed target-specific
intervention assignment, with \(\emptyset\) denoting no intervention
(numerically encoded as a binary indicator or quantitative dose), \(W_r\) contains
the recorded covariates and proxy measurements that directly enter this
mechanism, \(\varepsilon_r\) is
variable-specific noise, and \(f_r\) is an otherwise general structural
mechanism. Components of \(W\) outside \(W_r\) may still be used for
adjustment. We call a component a \emph{proxy measurement for unmeasured
confounding} when it carries information about the unmeasured confounders relevant
to a comparison; other components of \(W\) may be ordinary observed
covariates. The errors
\(\varepsilon_1,\ldots,\varepsilon_p\) are mutually independent and
independent of \((H,U,W)\), where \(U=(U_1,\ldots,U_p)\). Minimality means
that, after fixing the intervention assignment, recorded exogenous inputs, and unobserved
inputs, no proper subset of \(\Pa{r}\) yields the same structural mechanism
almost surely. Thus each \(j\in \Pa{r}\) makes a nonredundant direct
contribution to \(X_r\).

The recorded \(U_r\) is an intervention-assignment variable, such as a
guide indicator; it is distinct from the post-intervention value of \(X_r\).
Identification therefore compares observable conditional distributions across levels
{of the candidate assignment after conditioning on the remaining assignment design}
{\(U_{-r}\)}. When \(U_r\) shifts the distribution of
\(X_r\) and satisfies the conditional-exogeneity and target-specific exclusion
conditions below, it has the conventional role of an instrument for \(X_r\).

The assignments \(U\) and the recorded block \(W\) have complementary
roles. Source-specific intervention variation identifies whether \(X_j\) lies
upstream of \(X_r\). Within the resulting ancestor set, the same assignment
provides an intervention-assignment comparison, alongside the expression
comparison based on \(X_j\). PABM evaluates both through
conditional-distribution comparisons without an edge-specific coefficient
moment equation. The block \(W\) is {included in both fitted models}.
Its components may enter the target mechanism, track \(H\), or do both. In a
randomized perturbation study, \(W\) may contain baseline cell state, assay
controls, batch measurements, technical covariates, or measurements from
another modality.

{The structural equations define an acyclic observed-variable graph on
\([p]\), with edge set \(E=\{(j,r):j\in\Pa{r}\}\); its causal order may be
unknown. We write \(A\to B\) when \(A\) is a direct cause, or parent, of
\(B\); the arrow points from cause to effect. Thus graph recovery is equivalent
to recovering
\(\Pa{1},\ldots,\Pa{p}\). Hidden
confounding obstructs this task because a nonparent can remain highly
predictive of \(X_r\) by tracking \(H\). Graph recovery must therefore
distinguish a direct contribution to the mechanism for \(X_r\) from
predictability induced by the shared unmeasured confounders. Section~2.2 defines a
second target \(E^R\subseteq E\), the structural edges that respond to the
supported intervention design.}

The observable data are \((X,U,W)\). Only the components of \(X\) are
graph nodes. The assignment vector \(U\) supplies directional and
intervention-response information, while \(W\) is {included in the two
models fitted for each candidate--target pair}. The latent confounder \(H\) is never observed
or supplied to the estimator.
{Any component of \(W\) used in the ancestor screen must preserve the
conditional exogeneity in Assumption~\ref{ass:exogenous-u}(a); variables used
in a direct intervention-assignment comparison must support the separation in
Assumption~\ref{ass:proxy-sufficient}(b). Pre-intervention covariates are
natural candidates, whereas variables jointly affected by assignment and an
unmeasured confounder can violate these requirements.}

\subsection{Ancestor Identification and Proxy-Adjusted Parent Identification}

For each target \(X_r\), let
\(\mathcal C_r\subseteq[p]\setminus\{r\}\) denote its initial
candidate-parent set. Temporal information, perturbation design, scientific
knowledge, or an independent screening rule may restrict \(\mathcal C_r\);
without such information, take
\(\mathcal C_r=[p]\setminus\{r\}\). Full recovery of \(\Pa{r}\) requires
\emph{candidate coverage}, \(\Pa{r}\subseteq\mathcal C_r\). Without coverage,
the identifiable local target is the subset of parents retained in
\(\mathcal C_r\). {The finite-sample result for comparisons that adjust for the other
candidates while omitting the candidate under test additionally requires} \emph{ancestral
coverage}, \(\operatorname{Anc}(r)\subseteq\mathcal C_r\), so that every
intermediate ancestor {included in the population comparison is also
included in the sample comparison}.

{Candidate eligibility and adjustment have different purposes. A
scientific restriction may exclude a variable as a possible parent without
making it dispensable for adjustment. For example, along a mediated path
\(X_j\to X_k\to X_\ell\to X_r\), scientific knowledge may rule out
\(X_k\to X_r\) as a direct edge while \(X_k\) remains an ancestor that can
be retained in the adjustment set when assessing \(j\to r\). Candidate
restrictions alone therefore do not justify removing variables from adjustment. Our implementation uses the screened ancestor set for both purposes;
the theory therefore states candidate and ancestral coverage separately.}

We refine \(\mathcal C_r\) in two stages. The ancestor-screening stage compares
the conditional distribution of \(X_r\) with and without the source assignment \(U_j\).
Its expected conditional log-likelihood-ratio gain, equivalently a conditional
{Kullback--Leibler (KL) divergence, a measure of the difference between
conditional distributions, is positive exactly when} \(X_j\) is an ancestor
of \(X_r\) under the intervention conditions stated below.

\begin{samepage}
The parent-identification stage then separates direct parents from indirect
ancestors. It uses an expression comparison that reveals \(X_j\). When the
design supports the {candidate assignment contrast}, it also uses an intervention-assignment
comparison that reveals \(U_j\). Each comparison conditions on the other
ancestors to block indirect directed paths, adds the recorded covariate and
proxy block for adjustment, and retains the assignment terms required by its
nonedge argument. The maximum of the two direct-edge gains provides structural
evidence, while the intervention-assignment gain also records responsiveness
under the observed design.
\par
\end{samepage}

\begin{figure}[!tbp]
\centering
\includegraphics[width=0.98\textwidth]{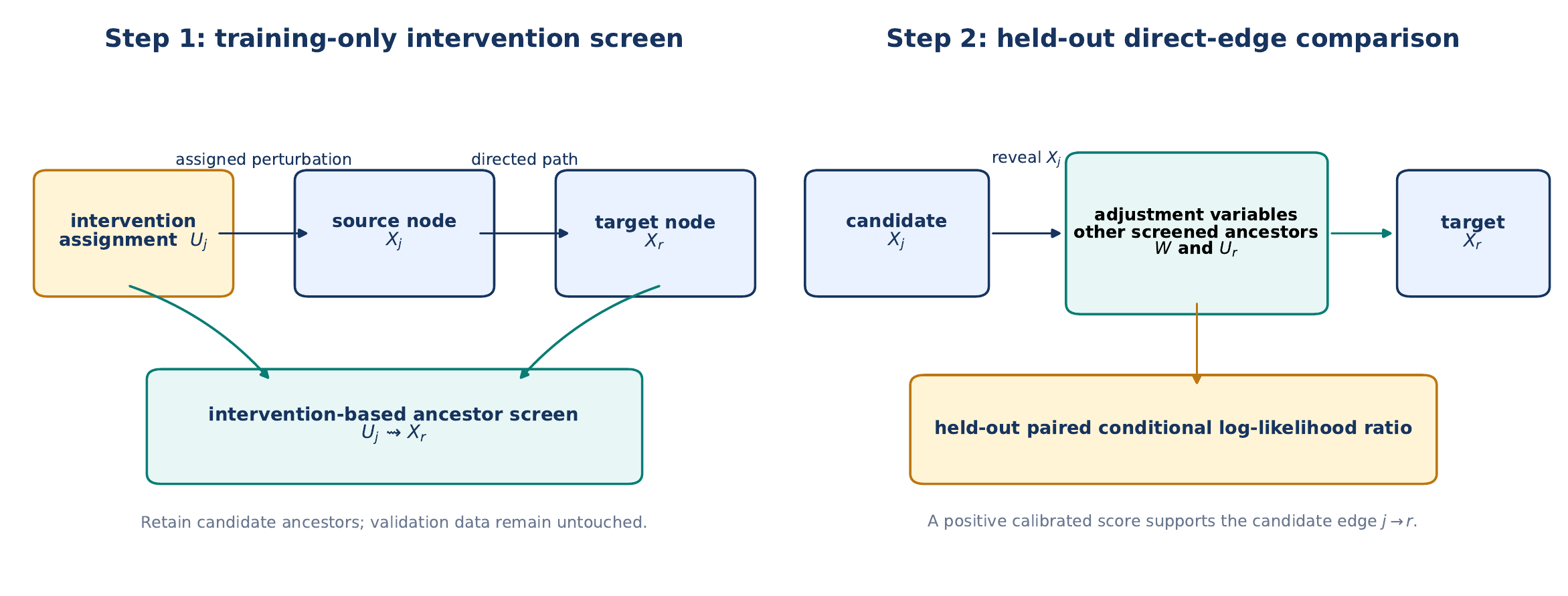}
\caption{{Two-stage PABM construction.} In the left panel, under
Assumption~\ref{ass:exogenous-u}, changing a possible source \(X_j\) through
its assigned intervention \(U_j\) reveals whether the target \(X_r\)
lies downstream, equivalently whether \(X_j\) is an ancestor of \(X_r\). This
training-only screen establishes ancestral eligibility; Stage~B evaluates the
direct edge \(X_j\to X_r\). In the right panel, for every screened
candidate, the held-out {conditional-likelihood comparison} evaluates whether \(X_j\)
contributes directly to \(X_r\) after the other training-selected candidates,
the recorded covariate and proxy block \(W\), and the target assignment \(U_r\)
are held
fixed. Identification requires the {conditional-independence requirements in}
Assumption~\ref{ass:proxy-sufficient}; a noisy proxy alone generally leaves an
unmeasured-confounding path open. The right panel depicts the expression
comparison. The intervention-assignment comparison analogously reveals the source
assignment when the design-support condition holds.}
\label{fig:assumption2-discriminability}
\end{figure}

\begin{samepage}
\noindent\textbf{Stage A: Intervention-based ancestor identification.}
For a fixed target \(X_r\), this step asks which source assignments \(U_j\)
change the conditional distribution of \(X_r\). Under the conditional exogeneity
assumption below, the observed-data comparison represents an intervention
contrast across supported levels of \(\operatorname{do}(U_j=u)\), conditional
on the remaining assignment design. Because \(U_j\) directly perturbs only
\(X_j\), any incremental association between \(U_j\) and \(X_r\) must
propagate along a directed path from \(X_j\) to \(X_r\). Under intervention
{validity, remaining assignment variation after conditioning (nondegenerate
support), and ancestor detectability, such an association is present exactly when}
\(X_j\) is a strict ancestor of \(X_r\).
\par
\end{samepage}

Formally, let \(U_{-j}\) denote the subvector of \(U\) excluding \(U_j\), and
omit \(W\) when no covariate or proxy block is recorded. Because the assignment
coordinates may be correlated, Stage~A measures the incremental information in
\(U_j\) after conditioning on \((U_{-j},W)\). Let
\(p_r(\,\cdot\mid\cdot)\) denote the true conditional density of \(X_r\).
For probability measures \(P\) and \(Q\), write
\(K(P,Q)=\int\log(dP/dQ)\,dP\) when \(P\) is absolutely continuous with
respect to \(Q\), and \(K(P,Q)=+\infty\) otherwise. For densities \(p\) and
\(q\) with respect to a common measure, this becomes
\(K(p,q)=\int\log(p/q)\,p\,d\mu\). All displayed gains are assumed finite.
For \(j\neq r\), define
\begin{equation}
\Delta^{\mathrm{anc}}_{jr}
:=
\mathbb E\!\left[
K\!\left(
p_r(\,\cdot\mid U,W),
p_r(\,\cdot\mid U_{-j},W)
\right)
\right].
\label{eq:ancestor-gain}
\end{equation}
The inner divergence integrates over \(X_r\) conditional on \((U,W)\); the
outer expectation is taken with respect to the joint distribution of
\((U,W)\). Thus \(\Delta^{\mathrm{anc}}_{jr}\) is the expected
conditional log-likelihood-ratio gain from adding \(U_j\) after conditioning
on \((U_{-j},W)\), and it equals zero precisely when adding \(U_j\) leaves the
conditional distribution of \(X_r\) unchanged almost surely. Define
\[
\mathcal A_r
:=
\{j\in[p]\setminus\{r\}:\Delta^{\mathrm{anc}}_{jr}>0\}.
\]

\begin{assumption}[Intervention validity, nondegenerate support, and ancestor detectability]
\label{ass:exogenous-u}
For the fixed target \(r\) and every possible intervention site \(j\neq r\),
the recorded assignments satisfy the following conditions.
\begin{enumerate}[label=(\alph*)]
\item \textbf{Conditional exogeneity and nondegenerate support.}
Given the remaining assignments \(U_{-j}\) and the recorded covariate
and proxy block \(W\), the source assignment is independent of the
unmeasured confounders,
\[
U_j\perp H\mid(U_{-j},W){.}
\]
Here and below, \(\perp\) denotes independence, conditional on the variables
following \(\mid\). The conditional distribution of \(U_j\) is nondegenerate
on a positive-probability
subset of the support. For a numerical encoding, a sufficient formulation is
\[
\Pbb\!\left\{\operatorname{Var}(U_j\mid U_{-j},W)>0\right\}>0.
\]
Here and below, conditional variance is taken over the displayed variable
{given the variables following the vertical bar, and the outer probability
is taken over those variables}.
This is a weak, identification-level support condition: a {candidate assignment contrast}
exists on a positive-probability set. {Stable estimation may require
sufficient conditional assignment variation throughout the relevant part of
the covariate distribution.}
Ancestor detectability in part (c) is required over this varying subset.

\item \textbf{Target-specific intervention and structural invariance
(exclusion restriction).}
Under \eqref{eq:structural-system}, \(U_j\) enters only the structural
equation for \(X_j\). For every \(k\neq j\), the mechanism \(f_k\) and parent
set \(\Pa{k}\) remain unchanged across values of \(U_j\). Consequently,
\(U_j\) can affect \(X_r\) only through a directed path from \(X_j\) to
\(X_r\).

\item \textbf{Pathwise relevance (ancestor detectability).}
Every strict ancestor of \(r\) produces detectable conditional intervention
variation at \(X_r\): for each \(j\in\operatorname{Anc}(r)\),
\(\Delta^{\mathrm{anc}}_{jr}>0\).
\end{enumerate}
\end{assumption}

Parts (a) and (b) imply \(\Delta^{\mathrm{anc}}_{jr}=0\) for every nonancestor of \(r\),
while part (c) gives \(\Delta^{\mathrm{anc}}_{jr}>0\) for every ancestor. Consequently,
$\mathcal A_r=\operatorname{Anc}(r)$.
Equivalently, \(j\in\operatorname{Anc}(r)\) if and only if
\(r\in\operatorname{Dec}(j)\); the ancestor screen is the transpose of the
graph-wide intervention--response relation. In a randomized perturbation
study, the assignment design can support conditional exogeneity and
nondegenerate assignment support. Target specificity, structural invariance, and
detectable propagation from every ancestor require additional scientific
justification.
A direct off-target action of \(U_j\) on another node can make \(X_r\)
respond for a nonancestor \(X_j\).

The eligible candidate-parent set is
$\mathcal C_r^A=\mathcal C_r\cap \mathcal A_r$.
Here the superscript (A) on \(\mathcal C_r^A\) denotes ancestor
eligibility.
Because every parent is a strict ancestor of its child in a DAG, exact
ancestor screening preserves every parent covered by \(\mathcal C_r\) while
removing descendants and causally unrelated nodes.

\noindent\textbf{Stage B: Proxy-adjusted direct-parent identification.}
After Stage~A, each \(j\in\mathcal C_r^A\) is a direct parent or an indirect
ancestor of \(X_r\). For \(I\subseteq[p]\), write
\(X_I=(X_k:k\in I)\). The other ancestors and \(W\) form the common
adjustment information. The block
\(X_{\mathcal A_r\setminus\{j\}}\) blocks every indirect
directed path from an indirect ancestor through another ancestor of \(r\),
whereas \(W\) may reduce association with unmeasured confounding.
{Every directed \(j\)-to-\(r\) path of length at least two has an internal
vertex in \(\mathcal A_r\setminus\{j\}\); conditioning on that vertex blocks
the path.}
Assumption~\ref{ass:proxy-sufficient} requires {the variables included
together} to establish the relevant conditional independence. The
intervention-assignment comparison conditions on \(U_{-j}\)
to control nonfocal intervention arms that may be associated with \(U_j\) and
may affect \(X_r\). {{For example, a one-hot single-guide experiment uses binary indicators
for mutually exclusive guide assignments, with all indicators zero for
controls. In this design,} the stratum \(U_{-j}=0\) leaves the
guide-\(j\)-versus-control contrast in \(U_j\); in every other stratum
\(U_j=0\).}
{The variables included in the reduced model for each comparison are}
\[
Z^X_{jr}=(X_{\mathcal A_r\setminus\{j\}},W,U_r),
\qquad
Z^U_{jr}=(X_{\mathcal A_r\setminus\{j\}},W,U_{-j}).
\]
{Write \(V_j^X=X_j\) and \(V_j^U=U_j\).
A simple graphical condition supplies the latent-adjusted independence
used below for an eligible nonparent \(j\notin\Pa{r}\). In an augmented causal DAG for \((X,U,W,H)\), suppose the
coordinates of \(Z^c_{jr}\) and \(V_j^c\) are nondescendants of \(X_r\),
and \(Z^c_{jr}\) together with \(H\) contains every parent of \(X_r\).
For \(c=U\), also require the target-specific exclusion
\(U_j\not\to X_r\). {The local Markov property states that a node is independent of its
nondescendants given its parents. Applied to the augmented DAG, it gives}
\(X_r\perp V_j^c\mid(Z^c_{jr},H)\). Thus blocking the mediated directed
paths is one part of the argument; admissibility of the additional recorded
variables and target specificity are also required.}
The expression comparison fixes the target assignment \(U_r\); the
intervention-assignment comparison fixes the nonfocal assignments \(U_{-j}\).
{Under exhaustive binary one-hot assignment without a control state,
\(U_j\) is determined by \(U_{-j}\), making the focal reveal unavailable.}

Define the two exact direct-edge gains
\begin{equation}
\Delta^X_{jr}=\E K\!\left\{p_r(\,\cdot\mid Z^X_{jr},X_j),p_r(\,\cdot\mid Z^X_{jr})\right\},
\qquad
\Delta^U_{jr}=\E K\!\left\{p_r(\,\cdot\mid Z^U_{jr},U_j),p_r(\,\cdot\mid Z^U_{jr})\right\}.
\label{eq:parent-gains}
\end{equation}
The first outer expectation is over the joint distribution of
\((Z^X_{jr},X_j)\), and the second is over that of \((Z^U_{jr},U_j)\); in
each case, the inner divergence integrates over the conditional distribution
of \(X_r\).
Both gains are nonnegative and are assumed finite. Each equals zero precisely
when the corresponding revealed variable is conditionally independent of
\(X_r\) {after the variables in \(Z^c_{jr}\) have been included}.

The nonedge conditions below can be written either as conditional
independences or as zero gains. The former state their causal content; the
latter provide regression targets for PABM's likelihood comparisons.

The intervention-assignment comparison is scientifically interpretable only
when the observed design contains a nondegenerate {candidate assignment contrast}. Record this
condition with the binary availability indicator $\mathbb{I}\{\cdot\}$:
\begin{equation}
a^U_{jr}
=\mathbb{I}\!\left\{
\Pbb\!\left\{\operatorname{Var}(U_j\mid Z^U_{jr})>0\right\}>0
\right\}.
\label{eq:u-availability}
\end{equation}
Here the conditional variance is taken under the distribution of \(U_j\)
given \(Z^U_{jr}\), and the outer probability is with respect to the marginal
distribution of \(Z^U_{jr}\).
Accordingly, \(a^U_{jr}=1\) precisely when a nondegenerate {candidate assignment contrast}
occurs {for values of \(Z^U_{jr}\) having positive probability}. {The
indicator records identification-level availability; stable finite-sample
estimation additionally requires sufficient conditional assignment variation.}
A pair with \(a^U_{jr}=0\) has an unavailable intervention-assignment
comparison. Its responsiveness status is unidentified under the observed
design.

{In computation, availability must be assessed with a prespecified support
diagnostic. {The ridge residual-variance rule used in Section~5 measures assignment
variation remaining after ridge-regression adjustment. It is only a
diagnostic:} positive linear residual variance need not imply positive
conditional variance under nonlinear dependence. For example, if
\(U_j=Z^2\) for \(Z\sim N(0,1)\), linear residualization on
\(Z\) leaves variation even though \(\operatorname{Var}(U_j\mid Z)=0\).
PABM omits the assignment comparison whenever its empirical support rule
fails, and then uses the expression comparison alone.}

Assumption~\ref{ass:proxy-sufficient} gives the nonedge null for each
conditional comparison. Part (a) says that \(X_j\) adds no information about
\(X_r\) given \(Z^X_{jr}\); when available, part (b) says the same for \(U_j\)
given \(Z^U_{jr}\). Conditioning on
\(X_{\mathcal A_r\setminus\{j\}}\) blocks indirect paths such
as \(X_j\to X_k\to X_r\), but may leave or induce association through \(H\).
{Sufficient conditions for the two nulls are}
\[
\begin{aligned}
X_j\perp X_r\mid(Z^X_{jr},H),\quad H\perp X_j\mid Z^X_{jr}
&\ \Longrightarrow\ X_j\perp X_r\mid Z^X_{jr}
&&\text{(expression)},\\
U_j\perp X_r\mid(Z^U_{jr},H),\quad H\perp U_j\mid Z^U_{jr}
&\ \Longrightarrow\ U_j\perp X_r\mid Z^U_{jr}
&&\text{(intervention assignment)}.
\end{aligned}
\]
In the expression line, structural blocking gives the first premise when the
other-ancestor block contains the parents of \(X_r\); {the second is independence from the latent causes after adjustment}.
In the intervention-assignment line, {the first premise requires
target-specific exclusion; \(U_{-j}\) can also remove association induced by
a dependent assignment design, {as required by the independence premise}. Whether a
nondegenerate {candidate assignment contrast} remains is recorded separately by \(a^U_{jr}\).}
{The second premise requires independence from the latent causes after adjustment.}
{Standard conditional-independence rules yield parts (a)--(b).} {These independence premises}
are additional identification requirements beyond proxy relevance and
randomization. Part (c) separately requires a true parent to have positive
gain in at least one available comparison.

\begin{assumption}[Proxy-adjusted nonedge separation and parent detectability]
\label{ass:proxy-sufficient}
For every \(j\in\mathcal C_r^A\), the following conditions hold.
\begin{enumerate}[label=(\alph*)]
\item \textbf{Expression nonedge separation.} If \(j\notin\Pa{r}\),
\[
X_j\perp X_r\mid Z^X_{jr},
\qquad\text{equivalently,}\qquad \Delta^X_{jr}=0.
\]
\item \textbf{Intervention-assignment nonedge separation.}
If \(j\notin\Pa{r}\) and
\(a^U_{jr}=1\),
\[
U_j\perp X_r\mid Z^U_{jr},
\qquad\text{equivalently,}\qquad \Delta^U_{jr}=0.
\]
\item \textbf{Parent detectability.} If \(j\in\Pa{r}\), then
\[
\Delta^X_{jr}>0
\quad\text{or}\quad
\{a^U_{jr}=1\ \text{and}\ \Delta^U_{jr}>0\}.
\]
\end{enumerate}
Parts (a)--(b) form the joint nonedge null, while part (c) allows either
comparison to detect a parent alternative.
\end{assumption}

{A constructive sufficient class clarifies the proxy requirement.
Suppose the target mechanism depends on \(H\) only through
\(B_r=b_r(H)\), and \(W\) determines \(B_r\). {The independent node-specific noise \(\varepsilon_r\) in
model~\eqref{eq:structural-system}} is independent of the ancestors,
assignments, \(W\), and \(H\); because \(Z^c_{jr}\) contains the target's
observed parents and required assignment terms, part (a) or (b) follows.
Latent variation unrelated to the target after \(B_r\) may remain unobserved.
Exact recording is essential for this class; an ordinary noisy proxy
generally yields approximate separation. Supplementary
Section~S2.1 gives the formal argument.}

\begin{remark}[Why Assumption~\ref{ass:proxy-sufficient} is needed and
how it can be relaxed]
\label{rem:a2-boundary}
{Causal discovery under unmeasured confounding requires an identifying
restriction that distinguishes direct edges from observationally equivalent
mediated or confounded associations. Indeed, if two causal structures induce
the same distribution for the recorded data but disagree on \(j\to r\), no
observed-data procedure can identify that edge. PABM expresses this requirement
through local, pair-specific conditional-distribution separation. Model-based
IV procedures encode identification through exclusion restrictions and
structural or measurement models; {neither formulation generally implies the other.}}

Once Stage~A has supplied the eligible ancestor family,
Assumption~\ref{ass:proxy-sufficient} is the exact separation condition for
PABM's population zero-versus-positive rule. Parts (a)--(b) give zero available
gain for every eligible nonparent, and part (c) gives positive gain in at least
one available comparison for every eligible parent. They are necessary and
sufficient for the score-level characterization in
Theorem~\ref{thm:population-identification}. Exact separation supplies the
population nulls used for identification and error control. {The approximate
theory requires the smallest parent gain to exceed the largest nonparent
gain by more than twice an upper bound on the largest score-estimation
error across comparisons;} see
Supplementary Corollaries~S9 and
S10.

PABM permits general conditional distributions for the target, while its
local separation conditions remain substantive, pair-specific, and generally
untestable from the observed distribution. They require {conditional independence beyond
proxy informativeness}. Proposition~\ref{prop:latent-imbalance} gives a sensitivity
interpretation of departures from exact separation.
\end{remark}

\begin{proposition}[{Residual latent-dependence bound}]
\label{prop:latent-imbalance}
Fix an eligible pair \((j,r)\) and a comparison \(c\in\{X,U\}\), where
\(V_j^X=X_j\) and \(V_j^U=U_j\). Suppose the structural and path-blocking
argument gives
\[
X_r\perp V_j^c\mid(Z_{jr}^c,H).
\]
Assume the required regular conditional distributions exist, and let
\(P_H(\,\cdot\mid\cdot)\) denote a conditional distribution of \(H\).
Assume the expected divergence below is finite, and define the {residual latent
dependence}
\[
\xi_{jr}^c=
\E\,K\!\left\{
P_H(\,\cdot\mid Z_{jr}^c,V_j^c),
P_H(\,\cdot\mid Z_{jr}^c)
\right\}.
\]
The expectation is with respect to the joint distribution of
\((Z_{jr}^c,V_j^c)\).
Then
\[
0\le \Delta_{jr}^c\le \xi_{jr}^c.
\]
Consequently, for an eligible nonparent, the blocking premise and
\(\xi_{jr}^c=0\) imply the corresponding nonedge condition in
Assumption~\ref{ass:proxy-sufficient}. For \(c=U\), the statement is needed
only when the intervention-assignment comparison is available.
\end{proposition}

{The bound gives a quantitative interpretation of proxy
adjustment: a spurious nonparent gain is no larger than the {latent dependence
remaining} after the recorded regression covariates have been fixed. {Exact latent
independence makes} this gain zero; {remaining latent dependence leaves} a positive upper bound.
Because \(H\) is unmeasured, \(\xi_{jr}^c\) is a sensitivity parameter.
It may be defined using a target-relevant summary of \(H\) whenever that
summary satisfies the blocking premise. A recorded block containing this
summary gives a simple {example of exact independence}. The Gaussian calculation in
Supplementary Section~S2.2 shows how
measurement noise can instead leave both nonparent gains positive. The proof
of the proposition is in Supplementary Section~S4;
Corollary~S10 connects its bound to recovery under
approximate separation.}

\begin{remark}[Complementary comparisons and size--power tradeoff]
For each eligible pair, PABM makes two separate comparisons. The expression
comparison conditions on \(Z^X_{jr}\) and reveals \(X_j\); the
intervention-assignment comparison conditions on
\(Z^U_{jr}\) and reveals \(U_j\).
{The two choices of \(Z_{jr}^c\) define separate conditional
likelihood-ratio comparisons}.
Residual variation in \(X_j\) may reveal a parent when assignment evidence is
weak or unavailable, and supported variation in \(U_j\) may reveal a
parent whose expression gain is weak after proxy adjustment.

{For an eligible nonparent, every available gain used must vanish; for a
parent, one positive available gain suffices. We call this the
joint-null/either-alternative structure.} Assumption
\ref{ass:proxy-sufficient}(a)--(b) supplies the joint nonedge null, and part
(c) supplies the either-comparison parent alternative. Theorem
\ref{thm:population-identification} establishes DAG identification from this
separation, and Theorem~\ref{thm:finite-dag-recovery} establishes finite-sample
recovery under uniform score accuracy and a positive population gap. The
edgewise complementarity and its single-comparison special cases are stated in
Corollary~\ref{cor:randomized}.
Comparative size and power remain design-dependent operating characteristics,
governed by signal strength, {conditional assignment variation}, {estimator accuracy}, proxy quality, and
dependence between the two statistics. {Adjusting for two comparisons can reduce power when assignment evidence is
weak or redundant; the combined p-value rule is given in Section~3.4.}
\end{remark}

{Returning to the example in the Introduction, we explain why both
nonparent nulls are required after ancestor adjustment. Suppose} the observed intermediate ancestor \(M\) satisfies
\[
U_j\to X_j\to M\to X_r,
\qquad H\to M,\quad H\to X_r,
\]
Each arrow points from a direct cause to its effect: \(M\) mediates the effect
of \(X_j\) on \(X_r\), while \(H\) is a common cause of \(M\) and \(X_r\).
Hence \(X_j\) is an indirect ancestor of
\(X_r\) through \(M\). Because
\(M\in\mathcal A_r\setminus\{j\}\), the
prescribed other-ancestor block conditions on \(M\) to block the indirect
directed path \(X_j\to M\to X_r\). The same conditioning opens the collider
path \(U_j\to X_j\to M\leftarrow H\to X_r\) {and opens the corresponding path}
\(X_j\to M\leftarrow H\to X_r\). Consequently, randomization of \(U_j\)
alone leaves a conditional association with \(X_r\) in the
intervention-assignment comparison, and \(X_j\) may remain conditionally
associated with \(X_r\) in the expression comparison. Part (b) imposes the
required zero-gain condition for the first comparison, and part (a) imposes it
for the second. An off-target path from \(U_j\) to \(X_r\) separately violates
the target-specific exclusion condition.

{The displayed sets of included variables define the population
comparisons used throughout the article.}

Throughout, pair subscripts are source-first and target-last: \(jr\) denotes
the candidate edge \(j\to r\). Using the {Stage~A} ancestor gain in
\eqref{eq:ancestor-gain}, define the population structural evidence by
\begin{equation}
\Omega_{jr}
=
\begin{cases}
\max\{\Delta^X_{jr},a^U_{jr}\Delta^U_{jr}\},
&j\in\mathcal C_r\ \text{and}\ \Delta^{\mathrm{anc}}_{jr}>0,\\
0,&\text{otherwise}.
\end{cases}
\label{eq:population-directed-weight}
\end{equation}
{The responsive set contains structural edges with positive assignment
gain; the concordant set contains structural edges with positive gains in
both the expression and assignment comparisons:}
\begin{equation}
E^R=\{(j,r)\in E:a^U_{jr}=1,\ \Delta^U_{jr}>0\},\qquad
E^C=\{(j,r)\in E:a^U_{jr}=1,\ \Delta^X_{jr}>0,\ \Delta^U_{jr}>0\}.
\label{eq:responsive-concordant-edge-sets}
\end{equation}
The superscript \(R\) refers to responsiveness under the recorded intervention
design. The set changes with intervention type, dose, and support. The
concordance set \(E^C\) {records positive gains in both comparisons; common}
violations of both nonedge conditions can still contaminate it.

The candidate-specific conditions should be tied to the data-generating
design. In a Perturb-seq or CRISPR screen, baseline cell-state measurements,
batch and quality-control covariates, surrogate factors, control features, and
other recorded modalities may contribute to \(W\). For the expression
comparison, this block must remove association between natural variation in
\(X_j\) and the unmeasured confounders relevant to \(X_r\). For the
intervention-assignment comparison, it must remove the corresponding
post-adjustment association while preserving {sufficient conditional variation} in the focal assignment.
Proxy ablations, intervention-assignment ablations,
negative-control analyses, {assignment-support diagnostics}, and split-sample stability
assess these requirements without proving them from observed data alone.

\subsection{Population Identification}

PABM's population identification argument decomposes graph recovery into five
tasks. Assumption~1 and candidate coverage
form the ancestor-eligible family; Assumption~2(a)--(b) separates eligible
nonparents in every included comparison; part (c) detects each parent in at
least one comparison; and the positive local gains determine the graph.
{Approximation within a chosen family of conditional-distribution models
(the working family), estimation, and graph calibration enter later} when these population quantities are estimated.

\begin{theorem}[Population identification by PABM]
\label{thm:population-identification}
Let \(r\in[p]\) be a target node. If the ancestor gains in
\eqref{eq:ancestor-gain} are well defined and finite and
Assumption~\ref{ass:exogenous-u} holds, then {Stage~A} identifies
\[
\mathcal A_r=\operatorname{Anc}(r),
\qquad
\mathcal C_r^A=\mathcal C_r\cap\operatorname{Anc}(r).
\]
If, in addition, the direct-edge gains in \eqref{eq:parent-gains} are well
defined and finite and Assumption~\ref{ass:proxy-sufficient} holds, then every
\(j\in\mathcal C_r^A\) satisfies
\[
j\in\Pa{r}
\quad\Longleftrightarrow\quad
\Omega_{jr}>0.
\]
If \(\Pa{r}\subseteq\mathcal C_r\), this equivalence identifies \(\Pa{r}\).
If these conditions and candidate coverage hold for every target, then
\[
E=\{(j,r):\Omega_{jr}>0,\ j\ne r\}.
\]
Within this structural graph, equation~\eqref{eq:responsive-concordant-edge-sets}
identifies the design-specific responsive and
concordant edge sets. Pairs with \(a^U_{jr}=0\) retain an unknown response
status.
\end{theorem}

The proof is in Supplementary Section~S4.

\begin{corollary}[Complementarity and randomized assignments]
\label{cor:randomized}
Under the theorem's nonedge conditions, PABM identifies every parent when
each parent has a positive gain in at least one available comparison. Different
parents may be detected by different comparisons. For example, one parent may
be expression-detectable and another intervention-detectable. A training rule
that omits the intervention component therefore requires positive expression
gain for every parent retained in the graph-recovery target.

This relaxation applies to {Stage~B} only. {Stage~A} still requires every structural
parent to survive the intervention-based ancestor screen. Both direct-edge
comparisons omit a parent removed by this screen. {Any broader empirical
candidate rule requires its own coverage and conditional-independence
analysis.}

If \(U\) is randomized independently of
\((H,W,\varepsilon_1,\ldots,\varepsilon_p)\), randomization supplies the
baseline exogeneity used by {Stage~A}. Assumption~\ref{ass:proxy-sufficient}(b)
additionally requires separation after conditioning on the other ancestors.
Mutually exclusive interventions also require a nondegenerate control
contrast: exhaustive one-hot coding without a control state makes \(U_j\)
deterministic given \(U_{-j}\).
\end{corollary}

{Numerically, retaining both comparisons lowers structural Hamming distance
(SHD), which counts missing, extra, and reversed edges, relative} to using the
expression comparison alone in the linear, nonadditive, and both count panels;
the training-only selection rule retains only the expression comparison in the cosine
panel ({Supplementary Table~S8} and
Section~\ref{subsec:count-main}).

\section{DAG Reconstruction by PABM}

Section~2 showed that the population score \(\Omega_{jr}\) separates parents
from eligible nonparents under Assumptions~1--2. PABM estimates this score with
a train--validation split. Training data determine the eligible pairs,
{the variables included in each comparison}, {comparison components selected for use, tuning choices, and}
fitted conditional distributions; {all selection and fitting decisions
therefore depend exclusively on the training sample. Validation data enter
only through evaluation of the fixed reduced--augmented comparisons and graph
calibration.} {The accepted directions are added in a prespecified order, skipping any
edge that would create a directed cycle. This step is called acyclic
projection. Retained edges are annotated by their intervention-response
evidence.}

Before computation, the analyst prespecifies the design restrictions, split,
{candidate-selection rules, variables included in each fitted model}, fitting procedures, component-selection and
pairwise-availability rules, {the calibration policy (the prespecified map from
the training component decision to a calibration method)}, and any familywise error
budgets.

\noindent\textbf{Algorithm 1 (PABM).}
The procedure has four computational phases.
\begin{enumerate}[label=\textbf{Phase \arabic*.}, leftmargin=2.8em]
\item \textbf{{Candidate selection}.} Determine the eligible ordered pairs and
{the variables to be included in their fitted models} from the training sample.
\item \textbf{{Model fitting}.} Select the active comparison components and
{fit each balanced pair of reduced and augmented conditional distributions.
Here active means selected for use in training; supported means passing the
empirical assignment-variation diagnostic; available refers to population
conditional assignment variation, as defined in Section~2.}
\item \textbf{{Validation}.} Compute held-out conditional
log-likelihood-ratio scores and apply the {calibration rule determined by
the training-stage component selection} to
obtain accepted structural directions and any prespecified responsive or
concordant directions.
\item \textbf{{DAG construction}.} Project the accepted structural directions onto
a DAG and intersect the resulting edge set with the responsive and concordant
decisions.
\end{enumerate}
{The required output is the estimated DAG. A prespecified implementation
may additionally report intervention-responsive and concordant annotations
when those families are computed.} The subsections below define the four
phases and their train--validation separation.

\subsection{\texorpdfstring{{Sample Splitting and Conditional-Distribution Estimation}}{Sample Splitting and Conditional-Distribution Estimation}}
\label{subsec:working-law-approximation}

For \(i\in[n]\), let \(X_i=(X_{i1},\ldots,X_{ip})\) and
\(U_i=(U_{i1},\ldots,U_{ip})\). Write
\(\mathcal D_n=\{(X_i,U_i,W_i):i\in[n]\}\), omitting \(W_i\) when no
recorded covariate or proxy block is available. The units are independent and
identically distributed, and the split is independent of their values. Let
\(I_{\mathrm{tr}}\dot\cup I_{\mathrm{val}}=[n]\), with sizes
\(n_{\mathrm{tr}}\) and \(n_{\mathrm{val}}\), and let
\(\mathcal D_{\mathrm{tr}}\) and \(\mathcal D_{\mathrm{val}}\) denote the
corresponding observations. Training {selects candidate pairs and the
variables included in each fit},
selects tuning parameters, and fits conditional distributions; validation is
reserved for the {held-out scores} and graph calibration. Let
\(\mathcal F_{\mathrm{tr}}\) {represent all information fixed by training, formally the sigma-field
generated by all training-stage}
objects, including the split, {selected candidates and included variables}, {empirical support
indicators, component-selection indicator},
tuning choices, working families, and fitted distributions.

For each comparison, the analyst chooses a conditional-distribution family
and a fitting procedure appropriate for the response. A Gaussian model with
a fitted mean, a Poisson regression, or a richer distributional regression
gives different approximations to the target distribution.
Section~\ref{subsec:pairwise-discovery} gives the fitted comparisons used in
the numerical studies. Section~4 shows how their approximation and estimation
errors enter graph recovery.

\subsection{\texorpdfstring{{Training-Stage Candidate Selection and Model Specification}}{Training-Stage Candidate Selection and Model Specification}}
\label{subsec:candidate-construction}

For target \(X_r\), let \(\mathcal J_r\subseteq[p]\setminus\{r\}\) contain
sources whose assignments can be used for intervention screening, and let
\(\mathcal C_r\subseteq[p]\setminus\{r\}\) contain variables scientifically
admissible as parents. These sets answer different questions and need not be
equal; without design, temporal, or scientific restrictions, both equal
\([p]\setminus\{r\}\).

\noindent\textbf{{Training-stage eligibility}.}
{Let
\(\widehat{\mathcal C}_r=\mathcal S_{C,r}(\mathcal D_{\mathrm{tr}};\eta_C)
\subseteq\mathcal C_r\) denote the output of a prespecified training-only
selection rule with fixed settings \(\eta_C\). The theoretical rule estimates
the population ancestor gain by a lower confidence bound (LCB) and intersects the
result with \(\mathcal C_r\); Supplementary
Proposition~S1 gives sufficient conditions.}
For \(j\in\widehat{\mathcal C}_r\), {collect the variables included in the
reduced fit for observation \(i\) as}
\[
\widehat Z^X_{ijr}=
(X_{i,\widehat{\mathcal C}_r\setminus\{j\}},
W_i,U_{ir}),\qquad
\widehat Z^U_{ijr}=
(X_{i,\widehat{\mathcal C}_r\setminus\{j\}},
W_i,U_{i,-j}).
\]
{Let
\(\widehat g_{\mathrm{tr}}=\mathcal S_g(\mathcal D_{\mathrm{tr}};\eta_g)
\in\{0,1\}\) be the prespecified training-only component-selection map, with
value one retaining the assignment comparison. For each eligible pair, let
\(\widehat a^U_{jr}=\mathcal S_{U,jr}(\mathcal D_{\mathrm{tr}};\eta_U)
\in\{0,1\}\) be its empirical assignment-support indicator. Their product
determines whether the assignment comparison is fitted.
Section~\ref{subsec:adaptive-comparison} summarizes the reported decision,
and Supplementary Section~S1.2 gives its exact
data-to-output map, including the pilot tests, ridge-support rule, thresholds,
and tested families.}

These training decisions fix the complete set of comparisons before validation
outcomes are used. Candidate selection has two statistical roles: retaining
the true parents and choosing the other responses used for adjustment.
Omitting an intermediate response may leave a mediated association, whereas
including an unsuitable response may induce an association through a
collider. Consequently, parent coverage and nonedge separation must both be
justified for the selected regression covariates. Sample splitting provides
independent evaluation of the resulting comparisons. Section~4 separately
states the coverage, separation, and assignment-support conditions needed to
interpret them as direct-edge evidence.

\subsection{Balanced Conditional-Likelihood Comparisons and Held-Out Scores}
\label{subsec:pairwise-discovery}

\begin{definition}[Balanced reduced--augmented comparison]
\label{def:balanced-masking}
For \(c\in\{X,U\}\), write \(V^X_j=X_j\), \(V^U_j=U_j\), and let
\(V^c_{ij}\) denote its value for unit \(i\). {Masking omits \(V^c_j\) from the reduced fit; the augmented fit adds it.} The comparison is balanced when
the two fits {use the same observations and target, include the same
variables \(\widehat Z^c_{jr}\), and use the same working distribution,
estimator, tuning rule, and split}. The augmented fit adds only \(V^c_j\).
{This comparability of the fitted models does not imply that adjustment
has removed unmeasured confounding.}
\end{definition}

\begin{figure}[!tbp]
\centering
\includegraphics[width=0.98\textwidth]{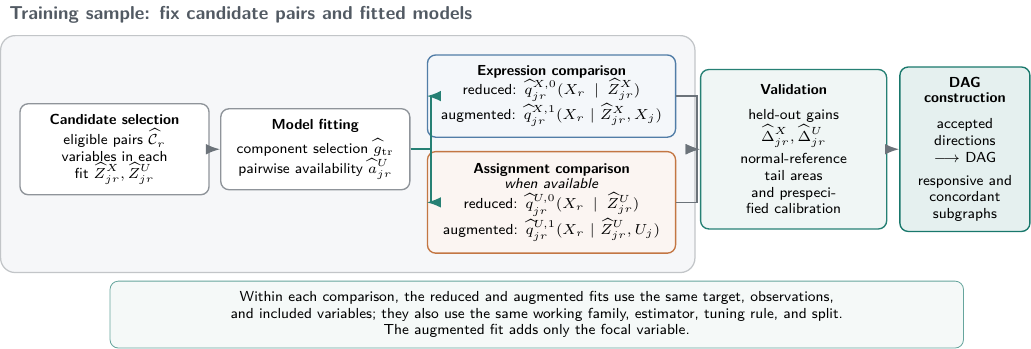}
\caption{PABM's train--validation construction for a candidate direction
\(j\to r\). Training data determine the eligible pair, {the variables
included in the two fitted models}, {the
comparisons selected using training data, and the balanced reduced and
augmented fits.} The expression
comparison adds \(X_j\); an available assignment comparison adds \(U_j\).
Validation data evaluate the fixed conditional log-likelihood ratios, {which
are used for testing, followed by removal of edges that would create a
directed cycle according to the prespecified ordering.} Within each comparison, the two fits
differ only by the candidate variable.}
\label{fig:pabm-hide-reveal}
\end{figure}

{For \(c\in\{X,U\}\), a prespecified fitting map
\(\mathcal L^c_{jr}\) returns the balanced pair of
conditional-distribution estimates}
\begin{equation}
{(\widehat q^{c,0}_{jr},\widehat q^{c,1}_{jr})
=\mathcal L^c_{jr}(\mathcal D_{\mathrm{tr}};
\widehat Z^c_{jr},V^c_j,\eta_L).}
\label{eq:fitted-direct-laws}
\end{equation}
The map specifies how the training data produce two conditional-distribution
estimates, including the working family, tuning, and optimization. For example,
in a unit-variance Gaussian model, estimate the reduced mean
\(\widehat m^{c,0}_{jr}(z)\) by least-squares regression on
\(\widehat Z^c_{jr}\), and estimate the augmented mean
\(\widehat m^{c,1}_{jr}(z,v)\) by the same procedure with \(V_j^c\) added.
Then \(\widehat q^{c,0}_{jr}(x\mid z)=
\phi\{x-\widehat m^{c,0}_{jr}(z)\}\) and
\(\widehat q^{c,1}_{jr}(x\mid z,v)=
\phi\{x-\widehat m^{c,1}_{jr}(z,v)\}\), where \(\phi\) is the standard
normal density. Section~\ref{subsec:gaussian-validation} uses ordinary least
squares. The subsequent benchmarks and K562 analysis estimate means by histogram
gradient boosting, which successively adds regression trees to reduce the
training loss \citep{friedman2001gradient,ke2017lightgbm}. For counts, the
same construction fits two Poisson means using the Poisson loss and the
{recorded exposure offset: \(\log L_i\) enters the log mean with coefficient
fixed at one, equivalently multiplying the fitted rate by \(L_i\).} Supplementary
Section~S1.2 specifies the offsets, constraints,
and tuning. Linear or sparse regression and neural-network fitting provide
other choices for the same construction.

The nonparametric structural framework leaves the conditional-distribution
estimator to be specified. A chosen working distribution may approximate
rather than equal the true conditional distribution and determines which
differences the fitted comparison can detect. The fitted Gaussian and Poisson models used here compare conditional
means. For a simple variance effect, let a binary parent \(X_j\) take values
zero and one with equal probability, and let
\(X_r=(1+X_j)\varepsilon_r\), where \(\varepsilon_r\sim N(0,1)\) is
independent of \(X_j\). The conditional mean is zero in both groups, while
the variance changes from one to four. With population-optimal fits, a
Gaussian mean-only comparison has zero likelihood gain; allowing the variance
to depend on the revealed variable gives a positive gain. Supplementary
Section~S2.4 derives both gains and the
relation between mean-estimation error and likelihood-gain error. This local
example isolates the choice of fitted distribution once a candidate is
eligible.

For \(i\in I_{\mathrm{val}}\), define
\begin{equation}
d^c_{ijr}=\log\frac{
\widehat q^{c,1}_{jr}(X_{ir}\mid\widehat Z^c_{ijr},V^c_{ij})}
{\widehat q^{c,0}_{jr}(X_{ir}\mid\widehat Z^c_{ijr})},
\qquad
\widehat\Delta^c_{jr}=\frac{1}{n_{\mathrm{val}}}
\sum_{i\in I_{\mathrm{val}}}d^c_{ijr}.
\label{eq:sample-direct-edge-lr}
\end{equation}
{We call the assignment component \emph{active} when
\(\widehat g_{\mathrm{tr}}=1\), an evaluated pair \emph{supported} when
\(\widehat a^U_{jr}=1\), and its population contrast \emph{available} when
\(a^U_{jr}=1\). An assignment score is evaluated only when the component is
active and the pair is supported.} For a standardized
continuous response under a unit-variance Gaussian working family,
\begin{equation}
\widehat\Delta^c_{jr}=\frac{1}{2n_{\mathrm{val}}}
\sum_{i\in I_{\mathrm{val}}}
\left[\{X_{ir}-\widehat m^{c,0}_{jr}(\widehat Z^c_{ijr})\}^2-
\{X_{ir}-\widehat m^{c,1}_{jr}(\widehat Z^c_{ijr},V^c_{ij})\}^2\right].
\label{eq:sample-direct-edge-gaussian}
\end{equation}
The Poisson implementation uses the analogous paired log-likelihood gain.

For score-separation analysis, define the raw structural score
\begin{equation}
\widehat\Omega_{jr}=
\begin{cases}
\max\{\widehat\Delta^X_{jr},\widehat\Delta^U_{jr}\},
&j\in\widehat{\mathcal C}_r,\ 
\widehat g_{\mathrm{tr}}\widehat a^U_{jr}=1,\\
\widehat\Delta^X_{jr},
&j\in\widehat{\mathcal C}_r,\ 
\widehat g_{\mathrm{tr}}\widehat a^U_{jr}=0,\\
0,&\text{otherwise}.
\end{cases}
\label{eq:estimated-directed-weight}
\end{equation}
This piecewise definition never multiplies an unevaluated score by zero and
preserves negative held-out evidence for an eligible pair.

\subsection{Calibration and Acyclic Projection}
\label{subsec:dag-discovery}
\label{subsec:pabm-estimation}

Let \(\overline d^c_{jr}\) and \(s^c_{jr}\) be the mean and sample standard
deviation of \(\{d^c_{ijr}:i\in I_{\mathrm{val}}\}\). The one-sided
normal-reference tail areas are
\[
T^c_{jr}=\frac{\overline d^c_{jr}}{s^c_{jr}/\sqrt{n_{\mathrm{val}}}},
\qquad \pi^c_{jr}=1-\Phi(T^c_{jr}).
\]
{Here \(\Phi\) is the standard normal cumulative distribution function.}
{If \(s^c_{jr}=0\), set \(\pi^c_{jr}=0\) when
\(\overline d^c_{jr}>0\), and set \(\pi^c_{jr}=1\) when
\(\overline d^c_{jr}\le 0\). {The value one for an exactly zero score is the nonrejecting convention
required for a valid null p-value. Such a p-value is called super-uniform:
conditional on training, its probability of being at most \(t\) is no
larger than \(t\), for every \(t\in[0,1]\).}}
The numerical implementation uses \(1/2\) for an exactly zero score.
Both conventions give identical reported rejections and graphs, because
neither value can pass the Benjamini--Hochberg (BH) or Holm procedures at
level 0.05 used here.
We refer to these tail
areas as working p-values when they enter a calibration rule; their inferential
validity requires the conditions in Section~4.
For each training-eligible pair, define
\begin{equation}
\pi^S_{jr}=\begin{cases}
\min\{1,2\min(\pi^X_{jr},\pi^U_{jr})\},
&\widehat g_{\mathrm{tr}}\widehat a^U_{jr}=1,\\
\pi^X_{jr},&\widehat g_{\mathrm{tr}}\widehat a^U_{jr}=0,
\end{cases}
\qquad
\pi^R_{jr}=\pi^U_{jr},\quad
\pi^C_{jr}=\max(\pi^X_{jr},\pi^U_{jr}),
\label{eq:dual-pvalues}
\end{equation}
where \(\pi^R_{jr}\) and \(\pi^C_{jr}\) are defined only when
\(\widehat g_{\mathrm{tr}}\widehat a^U_{jr}=1\). Their three roles are
\[
\begin{aligned}
S&:\ \text{structural evidence: at least one active component contributes},\\
R&:\ \text{intervention-responsive evidence: the assignment component contributes},\\
C&:\ \text{concordant evidence: both components contribute}.
\end{aligned}
\]
{The Bonferroni minimum permits structural evidence from either comparison,
while adjusting for the two tests. The maximum requires evidence from both
comparisons for concordance; it is an intersection--union test.}

The three {tested hypothesis families} are
\[
\mathcal H^S_{\mathrm{tr}}
=\{(j,r):j\in\widehat{\mathcal C}_r\},
\qquad
\mathcal H^R_{\mathrm{tr}}=\mathcal H^C_{\mathrm{tr}}
=\{(j,r)\in\mathcal H^S_{\mathrm{tr}}:
\widehat g_{\mathrm{tr}}\widehat a^U_{jr}=1\}.
\]
Conditional on \(\mathcal F_{\mathrm{tr}}\), these families and their sizes
\(M_F=|\mathcal H^F_{\mathrm{tr}}|\), \(F\in\{S,R,C\}\), are fixed.
{Under global Holm adjustment across all tested ordered pairs in each
relevant graph family, prespecify separate familywise error budgets}
\(\alpha_S,\alpha_R,\alpha_C\) and apply Holm at the corresponding level to
\(\{\pi^S_{jr}:(j,r)\in\mathcal H^S_{\mathrm{tr}}\}\),
\(\{\pi^R_{jr}:(j,r)\in\mathcal H^R_{\mathrm{tr}}\}\), and
\(\{\pi^C_{jr}:(j,r)\in\mathcal H^C_{\mathrm{tr}}\}\), producing rejection
sets \(\widehat{\mathcal R}^S\), \(\widehat{\mathcal R}^R\), and
\(\widehat{\mathcal R}^C\). If family \(F\in\{S,R,C\}\) contains \(M_F\)
tests, Holm orders its p-values and compares the \(k\)th smallest with
\(\alpha_F/(M_F-k+1)\)
\citep{holm1979simple}. {The calibration route is fixed by the training
component decision: \(\widehat g_{\mathrm{tr}}=0\) applies
BH \citep{benjamini1995controlling} separately to
\(\{\pi^X_{jr}:j\in\widehat{\mathcal C}_r\}\) for each target \(r\), whereas
\(\widehat g_{\mathrm{tr}}=1\) always applies global Holm to the structural
family and, when prespecified as outputs, separately applies Holm to the
responsive and concordance families. The reported continuous dual-route
implementation computes all three families; the count study reports only the
structural graph.}
Section~4 does not claim FWER control for the empirical targetwise-BH route.

{For graph-level error control, an analyst uses global Holm for the
structural family, with either the expression comparison alone or both
supported comparisons. This requires valid p-values for the comparisons
selected in training; multiplicity adjustment cannot repair a failed nonedge
restriction. The empirical expression-only implementation instead uses
targetwise BH and carries no graph-level FWER claim. Table~\ref{tab:guarantee-scope}
and Supplementary Section~S1.2 distinguish these
procedures and their fitting details.}

Let \(\widehat{\mathcal E}^S\) denote the structural directions accepted by
the {method realized under the prespecified calibration policy}. Thus
\(\widehat{\mathcal E}^S=\widehat{\mathcal R}^S\) under global Holm; a
score threshold, lower confidence bound, or targetwise BH rule supplies its
own accepted set. {The projection ranking is prespecified with the
calibration route. In the reported implementation, the continuous
expression-only route ranks BH-accepted pairs by decreasing raw gain, the
count expression-only route ranks by decreasing normal-reference statistic,
and the dual route ranks by increasing combined p-value. Supplementary
Section~S1.1.2 records the branch-specific
tie conventions used in the numerical implementation.}

Let \(K_S=|\widehat{\mathcal E}^S|\), and write the ranked directions as
\(e_1,\ldots,e_{K_S}\).
Starting from \(\widehat E^{\mathrm{DAG}}_0=\varnothing\), set
\begin{equation}
\widehat E^{\mathrm{DAG}}_\ell=
\begin{cases}
\widehat E^{\mathrm{DAG}}_{\ell-1}\cup\{e_\ell\},
&\text{if the union is acyclic},\\
\widehat E^{\mathrm{DAG}}_{\ell-1},&\text{otherwise},
\end{cases}
\quad
(\widehat\Gamma)_{jr}=\mathbb I\{j\to r\in
\widehat E^{\mathrm{DAG}}_{K_S}\}.
\label{eq:pabm-graph-estimator}
\end{equation}
Write \(\widehat E^{\mathrm{DAG}}=\widehat E^{\mathrm{DAG}}_{K_S}\) and
define the intervention-responsive and concordant subgraphs by
\[
\widehat E^R=\widehat E^{\mathrm{DAG}}\cap\widehat{\mathcal R}^R,
\qquad
\widehat E^C=\widehat E^{\mathrm{DAG}}\cap\widehat{\mathcal R}^C,
\]
when those families are included in the prespecified calibration {policy}.
The responsive and concordant families do not add structural directions;
their rejections label structurally retained edges through these intersections.
{When the assignment component is inactive, or active but the pair lacks
adequate training support,} responsiveness is not classified; nonrejection of an evaluated comparison
records a failure to detect a response.

Acyclic projection is an output constraint shared by all calibration methods.
Under exact recovery it is inert because the accepted set already equals the
true DAG. With selection errors, projection enforces acyclicity and may remove
directions according to their ranking; it has no separate identification
claim.

The structural, responsive, and concordance rejection sets are formal outputs
of separate tested families. Their intersections may be reported as descriptive
evidence patterns. An assignment-only pattern may indicate
off-target effects, collider bias, or loss of expression-comparison variation;
{an inactive component or unsupported pair is not evidence of nonresponsiveness,} and
population unavailability leaves responsiveness unidentified under the
observed design.

At most twice \(\sum_r|\widehat{\mathcal C}_r|\) balanced comparisons are
fitted. {Each comparison ordinarily requires a reduced and an augmented
fit, whose cost depends on the dimension of the selected regression
covariates. The comparison count alone therefore does not imply low fitting
cost: a sparse DAG may have large ancestor sets, and an assignment comparison
can contain \(p-1\) nonfocal assignments.} Scientific restrictions or a known
order can reduce this number; otherwise graph assembly performs one cycle
check per accepted structural {direction.}

\section{Statistical Guarantees for Graph Recovery}
\label{sec:pabm-theory}

The statistical question is how accurately the fitted conditional-likelihood
gains must approximate their population counterparts to recover the graph.
Proposition~\ref{prop:training-selection-recovery} answers this question for
a training-selected family that contains every true edge.
Theorem~\ref{thm:finite-dag-recovery} then expresses the required accuracy in
terms of working-model approximation, estimation error, and held-out
variability when training recovers the population ancestor sets and assignment
availability. Together with Proposition~\ref{prop:latent-imbalance}, these
results distinguish the effects of incomplete proxy adjustment from those of
estimating the conditional distributions. Theorem~\ref{thm:dual-fwer}
addresses the separate question of simultaneous false-positive control under
valid component p-values. Both conclusions depend on the comparisons selected
from training: recovery requires a sufficient parent--nonparent score gap,
whereas error control requires valid nonedge p-values for the selected
regression covariates. The results below state these requirements conditionally
and show how training-stage failure enters the unconditional guarantees.

Table~\ref{tab:guarantee-scope} connects the population construction,
training-stage choices, and numerical implementation. Applying a recovery
result requires coverage of true parents, separation under the selected
regression covariates, and accurate fitted gains. Simultaneous error control
additionally requires valid p-values for those comparisons. These requirements
can be studied separately for a chosen conditional-distribution estimator and
training-stage selection rule.

\begin{table}[t]
\centering
\caption{Scope of PABM's identification, recovery, and testing results.
All rows use the same pairwise comparison construction.}
\label{tab:guarantee-scope}
\small
\begin{tabularx}{\textwidth}{@{}>{\raggedright\arraybackslash}p{0.18\textwidth}>{\raggedright\arraybackslash}X>{\raggedright\arraybackslash}X@{}}
\toprule
Analysis & Candidate and decision requirements & Conclusion \\
\midrule
Population target & Population ancestors and prescribed regression covariates;
Assumptions~1--2; positive gains & Direct-parent identification
(Theorem~\ref{thm:population-identification}). \\
Realized training choices & Parent coverage; a parent--nonparent gain gap under
the selected covariates; uniform score accuracy; a separating threshold &
Conditional graph-recovery implication
(Proposition~\ref{prop:training-selection-recovery}) and unconditional transfer
after accounting for training-stage failure
(Corollary~\ref{cor:selected-context-transfer}). \\
Exact training-stage selection & Population candidate sets, regression covariates,
and active support indicators recovered; fitting-risk and validation bounds &
Explicit recovery probability
(Theorem~\ref{thm:finite-dag-recovery}); supplementary lower-confidence-bound
rules give one sufficient selection route. \\
{Inferential implementation} & {Training-fixed candidates and fits;
valid held-out p-values; global Holm, using expression alone or both supported
comparisons} & {Conditional graph-level FWER
(Theorem~\ref{thm:dual-fwer}); exact Gaussian least-squares calibration is one
finite-sample instance.} \\
{Numerical implementation} & {{Candidates from the empirical residual-association screen or its
replacement rule;}
boosting fits; working normal p-values; targetwise BH or global Holm selected
in training} & {Empirical recovery evidence (Section~5). The stated coverage
and calibration conditions remain to be established for these choices.} \\
\bottomrule
\end{tabularx}
\vspace{3pt}
\begin{minipage}{\textwidth}\footnotesize
{The empirical expression-only route uses targetwise BH and therefore has
no graph-level FWER guarantee.} Acyclic projection preserves exact recovery
and can only delete accepted directions.
\end{minipage}
\end{table}

Let \(\Gamma^\circ\) denote the true adjacency matrix under the source-first
convention, \(\Gamma^\circ_{jr}=\mathbb I\{j\in\Pa r\}\).
The technical screening, {estimator-risk}, concentration, robustness, and
additional {training-stage selection} results appear in Supplementary
Section~S3; all proofs are in Supplementary
Section~S4.

\subsection{\texorpdfstring{{Graph Recovery after Training-Stage Selection}}{Graph Recovery after Training-Stage Selection}}

{The training sample selects the candidate pairs, the variables included
in each reduced and augmented fit, and whether the assignment comparison is
used. Conditional on \(\mathcal F_{\mathrm{tr}}\), all these choices are
fixed.} Let
\(\widehat{\mathcal K}=\{(j,r):j\in\widehat{\mathcal C}_r\}\). {For an
independent observation from the study population, let \(\widehat Z^c_{jr}\)
collect the coordinates selected by training. For each selected pair define}
\begin{align}
{\Delta^{c,\mathrm{sel}}_{jr}}
&{=\E\!\left[
K\!\left\{
p_r(\,\cdot\mid\widehat Z^c_{jr},V_j^c,\mathcal F_{\mathrm{tr}}),
p_r(\,\cdot\mid\widehat Z^c_{jr},\mathcal F_{\mathrm{tr}})
\right\}\middle|\mathcal F_{\mathrm{tr}}\right],}\notag\\
{a^{U,\mathrm{sel}}_{jr}}
&{=\mathbb I\!\left[
\Pbb\!\left\{\operatorname{Var}(U_j\mid\widehat Z^U_{jr},
\mathcal F_{\mathrm{tr}})>0\middle|\mathcal F_{\mathrm{tr}}\right\}>0
\right].}
\label{eq:selected-population-quantities}
\end{align}
{The expectation and probability are over this fresh observation,
conditional on the realized training selection; \(p_r\) denotes its true
conditional distribution.} Define
\[
\Omega^{\mathrm{sel}}_{jr}
=\max\!\left\{\Delta^{X,\mathrm{sel}}_{jr},
\widehat g_{\mathrm{tr}}a^{U,\mathrm{sel}}_{jr}
\Delta^{U,\mathrm{sel}}_{jr}\right\}.
\]

\begin{proposition}[{Recovery after training-stage selection}]
\label{prop:training-selection-recovery}
Let \(E\) be the true edge set defined in Section~2. Suppose
\(E\subseteq\widehat{\mathcal K}\) and
{when \(\widehat g_{\mathrm{tr}}=1\),
\(\widehat a^U_{jr}=a^{U,\mathrm{sel}}_{jr}\) for every selected pair.} Let
\(B_{\mathrm{sel}}\) be a nonnegative, training-measurable radius satisfying
\[
\Pbb\!\left\{
\max_{(j,r)\in\widehat{\mathcal K}}
|\widehat\Omega_{jr}-\Omega^{\mathrm{sel}}_{jr}|
\le B_{\mathrm{sel}}\ \middle|\ \mathcal F_{\mathrm{tr}}\right\}
\ge 1-\delta_{\mathrm{sel}}.
\]
Define
\[
\eta_{\mathrm{sel}}
=\max_{(j,r)\in\widehat{\mathcal K}\setminus E}
\Omega^{\mathrm{sel}}_{jr},
\qquad
\kappa_{\mathrm{sel}}
=\min_{(j,r)\in E}\Omega^{\mathrm{sel}}_{jr},
\]
where \(\max\varnothing=-\infty\) and \(\min\varnothing=+\infty\). If
\(\kappa_{\mathrm{sel}}-\eta_{\mathrm{sel}}>2B_{\mathrm{sel}}\), then every
threshold satisfying
\[
\eta_{\mathrm{sel}}+B_{\mathrm{sel}}<\lambda_S
<\kappa_{\mathrm{sel}}-B_{\mathrm{sel}}
\]
gives
\[
\{(j,r)\in\widehat{\mathcal K}:
\widehat\Omega_{jr}>\lambda_S\}=E
\]
on the uniform-accuracy event. Greedy acyclic projection
then returns \(\Gamma^\circ\). Thus the conditional probability of exact
recovery is at least \(1-\delta_{\mathrm{sel}}\) whenever the displayed gap
and threshold conditions hold for the realized training sample.
\end{proposition}

\begin{corollary}[Unconditional recovery after training-stage selection]
\label{cor:selected-context-transfer}
Let \(\mathcal G_{\mathrm{rec}}\) be the training-measurable event on which
the coverage and availability conditions of
Proposition~\ref{prop:training-selection-recovery} hold, the threshold
\(\lambda_S\) is fixed or \(\mathcal F_{\mathrm{tr}}\)-measurable, and
\[
\kappa_{\mathrm{sel}}-\eta_{\mathrm{sel}}>2B_{\mathrm{sel}},
\qquad
\eta_{\mathrm{sel}}+B_{\mathrm{sel}}<\lambda_S
<\kappa_{\mathrm{sel}}-B_{\mathrm{sel}}.
\]
For deterministic \(\delta_{\mathrm{sel}},\delta_{\mathrm{rec}}\in[0,1]\),
suppose the proposition's conditional score-accuracy probability is at least
\(1-\delta_{\mathrm{sel}}\) for almost every training realization in
\(\mathcal G_{\mathrm{rec}}\), and
\(\Pbb(\mathcal G_{\mathrm{rec}}^c)\le\delta_{\mathrm{rec}}\). Then
\[
\Pbb(\widehat E_{\lambda_S}\ne E)
\le \delta_{\mathrm{rec}}
+(1-\delta_{\mathrm{rec}})\delta_{\mathrm{sel}}
\le \delta_{\mathrm{rec}}+\delta_{\mathrm{sel}}.
\]
The same bound applies after greedy acyclic projection.
\end{corollary}

{The proposition and corollary separate four requirements: the selected family covers
the true edges; empirical and population assignment availability agree when
that comparison is active; the population parent--nonparent gap is positive;
and the score error is smaller than half that gap. The quantity
\(\delta_{\mathrm{rec}}\) bounds the probability that training fails at least one
of these requirements.} Assumption~2 makes the nonparent term zero
{when the population candidate and model variables are used}. {The
proposition is a post-selection score-separation implication; it supplies no
validity result for any particular empirical selector. A useful bound on
\(\delta_{\mathrm{rec}}\) therefore requires a separate analysis of that
selector.} The proofs are in
Supplementary Section~S4.

\subsection{\texorpdfstring{{Explicit Score Accuracy under Exact Training-Stage Selection}}{Explicit Score Accuracy under Exact Training-Stage Selection}}

{Define the theoretical sets of comparisons using the true ancestor sets
and population assignment availability. We call these oracle sets because
they use population information unavailable to the estimator:}
\[
\mathcal K_X=\{(j,r):j\in\mathcal C_r\cap\operatorname{Anc}(r)\},
\qquad
\mathcal K_U=\{(j,r)\in\mathcal K_X:a^U_{jr}=1\},
\]
with sizes \(M_X\) and \(M_U\), each at most \(p(p-1)\). The {exact-selection}
analysis uses the good training event
\begin{equation}
\begin{aligned}
\mathcal G_{\mathrm{tr}}
&=\left\{\widehat{\mathcal C}_{r}
=\mathcal C_r\cap\operatorname{Anc}(r)\ \text{for every }r\right\}\\
&\quad\cap\left[\{\widehat g_{\mathrm{tr}}=0\}
\cup\left\{\widehat g_{\mathrm{tr}}=1,\
\widehat a^U_{jr}=a^U_{jr}
\ \text{for every }(j,r)\in\mathcal K_X\right\}\right].
\end{aligned}
\label{eq:good-training-event}
\end{equation}
All maxima over an empty comparison library are defined as zero; the
corresponding score-error requirement is then vacuous.
{On this event and under ancestral coverage,} every held-out comparison uses its {population candidate
set and the corresponding other-ancestor variables}. When the assignment component is active, empirical
training support also agrees with population availability.

For \(c\in\{X,U\}\), the four score levels are: the true gain
\(\Delta^c_{jr}\), {the gain \(\Delta^{c,\mathrm{proj}}_{jr}\) obtained from the best KL
approximations to the reduced and augmented conditional distributions
within the chosen working families}, the conditional mean of the fitted score
\(\overline\Delta^{c,\mathrm{fit}}_{jr}\), and the proof-only held-out score
\(\widehat\Delta^{c,\mathrm{orc}}_{jr}\) evaluated {with the population
variables \(Z^c_{jr}\)}.
They satisfy the exact decomposition
\begin{equation}
\widehat\Delta^{c,\mathrm{orc}}_{jr}-\Delta^c_{jr}
=\underbrace{\Delta^{c,\mathrm{proj}}_{jr}-\Delta^c_{jr}}_{\text{working-family approximation}}
+\underbrace{\overline\Delta^{c,\mathrm{fit}}_{jr}-
\Delta^{c,\mathrm{proj}}_{jr}}_{\text{{estimator fitting}}}
+\underbrace{\widehat\Delta^{c,\mathrm{orc}}_{jr}-
\overline\Delta^{c,\mathrm{fit}}_{jr}}_{\text{held-out evaluation}}.
\label{eq:score-error-decomposition}
\end{equation}
Here
\(\overline\Delta^{c,\mathrm{fit}}_{jr}
=\E(\widehat\Delta^{c,\mathrm{orc}}_{jr}\mid\mathcal F_{\mathrm{tr}})\).
Let \(\varepsilon^{\mathrm{app}}_c\) uniformly bound the first term over
\(\mathcal K_c\) and all selectable working families. For \(\delta\in(0,1)\),
let the deterministic estimator-specific fitting-error radius satisfy
\[
\Pbb\!\left\{\max_{(j,r)\in\mathcal K_c}
|\overline\Delta^{c,\mathrm{fit}}_{jr}-
\Delta^{c,\mathrm{proj}}_{jr}|
\le\rho_{c,n_{\mathrm{tr}}}(\delta)\right\}\ge1-\delta.
\]
Conditional on \(\mathcal F_{\mathrm{tr}}\), suppose the independent centered
{validation contributions obey a Bernstein tail-probability bound for their
average, with deterministic parameters}
\(\nu_c,b_c>0\). Allocating \(\delta_c/2\) to {estimator fitting} and
\(\delta_c/2\) to held-out evaluation, define
\begin{align}
V_{c,n_{\mathrm{val}}}(\delta_c)
&=\nu_c\sqrt{\frac{2\log\{4\max(1,M_c)/\delta_c\}}{n_{\mathrm{val}}}}
+\frac{b_c\log\{4\max(1,M_c)/\delta_c\}}{n_{\mathrm{val}}},\notag\\
B_c(\delta_c)
&=\underbrace{\varepsilon^{\mathrm{app}}_c}_{\text{approximation}}
+\underbrace{\rho_{c,n_{\mathrm{tr}}}(\delta_c/2)}_{\text{fitting}}
+\underbrace{V_{c,n_{\mathrm{val}}}(\delta_c)}_{\text{validation}}.
\label{eq:component-score-radius}
\end{align}
Only the last component has the generic split-sample order
\(\sqrt{\log(M_c)/n_{\mathrm{val}}}\); the fitting rate is supplied by the
chosen {estimator}, and approximation error may persist under misspecification.

For the {training-sample component-selection indicator}, define
\[
\Omega^{\mathrm{tr}}_{jr}=\begin{cases}
\max\{\Delta^X_{jr},\Delta^U_{jr}\},
&\widehat g_{\mathrm{tr}}a^U_{jr}=1,\\
\Delta^X_{jr},&\widehat g_{\mathrm{tr}}a^U_{jr}=0,
\end{cases}
\qquad
B_S=\begin{cases}
\max\{B_X(\delta_X),B_U(\delta_U)\},&\widehat g_{\mathrm{tr}}=1,\\
B_X(\delta_X),&\widehat g_{\mathrm{tr}}=0,
\end{cases}
\]
on the oracle candidate family, and set the score to zero otherwise. If the
assignment component is inactive, its accuracy requirement is vacuous. Because
the {component-selection indicator} is training-measurable,
\(\Omega^{\mathrm{tr}}_{jr}\), \(B_S\), and
the minimum active-edge signal below are training-random. For a nonempty true
edge set, ancestral coverage gives \(E\subseteq\mathcal K_X\), and we define
\(\kappa_S^{\mathrm{tr}}=\min_{(j,r)\in E}\Omega^{\mathrm{tr}}_{jr}\).
For the assignment component, define the active-score accuracy event
\[
\mathcal G_U^{\mathrm{act}}
=\{\widehat g_{\mathrm{tr}}=0\}
\cup\left\{\widehat g_{\mathrm{tr}}=1,
\max_{(j,r)\in\mathcal K_U}
|\widehat\Delta^{U,\mathrm{orc}}_{jr}-\Delta^U_{jr}|
\le B_U(\delta_U)\right\}.
\]
It holds automatically on every training realization with
\(\widehat g_{\mathrm{tr}}=0\). If the assignment component is identically
inactive, remove the {assignment component}, use \(B_S=B_X(\delta_X)\), and set its
failure contribution to zero; \(B_U\) then need not be defined. Otherwise,
\((\mathcal G_U^{\mathrm{act}})^c\) is precisely the intersection of the
{event that the assignment component is retained} and failure of the U-score bound.

For a fixed or \(\mathcal F_{\mathrm{tr}}\)-measurable threshold
\(\lambda_S\), write
\[
\widehat E_{\lambda_S}
=\{j\to r:j\in\widehat{\mathcal C}_r,\
\widehat\Omega_{jr}>\lambda_S\}
\]
and define the joint {selection-and-score} accuracy event
\[
\mathcal A_S=
\mathcal G_{\mathrm{tr}}\cap\left\{
\max_r\max_{j\in\mathcal C_r\cap\operatorname{Anc}(r)}
|\widehat\Omega_{jr}-\Omega^{\mathrm{tr}}_{jr}|\le B_S
\right\}.
\]

\begin{theorem}[Finite-sample PABM graph recovery]
\label{thm:finite-dag-recovery}
Suppose Assumptions~\ref{ass:exogenous-u} and
\ref{ass:proxy-sufficient} hold, each \(\mathcal C_r\) has ancestral coverage,
and every ancestor belongs to \(\mathcal J_r\). Assume
\(\Pbb(\mathcal G_{\mathrm{tr}})\ge1-\delta_{\mathrm{tr}}\), that the
uniform expression-score event with radius \(B_X(\delta_X)\) in
\eqref{eq:component-score-radius} has failure probability at most
\(\delta_X\), and that
\(\Pbb\{(\mathcal G_U^{\mathrm{act}})^c\}\le\delta_U\). Then
\[
\Pbb(\mathcal A_S^c)
\le\delta_{\mathrm{tr}}+\delta_X+\delta_U.
\]
On \(\mathcal A_S\), if \(E\ne\varnothing\),
\(2B_S<\kappa_S^{\mathrm{tr}}\), and
\[
B_S<\lambda_S<\kappa_S^{\mathrm{tr}}-B_S,
\]
then \(\widehat E_{\lambda_S}=E\), and ranked greedy
projection returns \(\Gamma^\circ\). If \(E=\varnothing\), any
\(\lambda_S>B_S\) recovers the empty graph on \(\mathcal A_S\).
\end{theorem}

Under exact selection, the projection is inert because the recovered edge set
is already the true DAG. For nonempty \(E\), the pointwise implication gives
\[
\begin{aligned}
\Pbb\{\widehat E_{\lambda_S}\ne E\}
&\le \Pbb(\mathcal A_S^c)
+\Pbb\!\left(\mathcal A_S\cap
\{\lambda_S\le B_S\ \text{or}\
\lambda_S\ge\kappa_S^{\mathrm{tr}}-B_S\}\right)\\
&\le\delta_{\mathrm{tr}}+\delta_X+\delta_U
+\Pbb\!\left\{\lambda_S\le B_S\ \text{or}\
\lambda_S\ge\kappa_S^{\mathrm{tr}}-B_S\right\}.
\end{aligned}
\]
Thus, if the separating inequalities hold with probability at least
\(1-\delta_S\) for some \(\delta_S\in[0,1]\), the graph-recovery error probability is at most
\(\delta_{\mathrm{tr}}+\delta_X+\delta_U+\delta_S\).

\noindent\textbf{Interpretation of the recovery bound.}
The terms on the right-hand side have different roles. The levels
\(\delta_X\) and \(\delta_U\) are prespecified confidence allocations for the
two component-score bounds. Their associated radii become explicit after
valid bounds are supplied for working-family approximation, {estimation error},
and validation tails. The quantity \(\delta_{\mathrm{tr}}\) is a
repeated-sampling upper bound on failure of the training-stage screen,
{the variables included in each fit}, and active support classifications. For the
{theoretical LCB screen},
Supplementary Proposition~S1 and a separate
design argument provide its screening and support components. The threshold
\(\lambda_S\) is fixed or selected from training data, whereas
\(\kappa_S^{\mathrm{tr}}\) is a minimum population edge gain and is generally
unknown. The final probability is therefore a signal--threshold separation
term. The display is a sufficient repeated-sampling recovery guarantee; a
data-computable certificate would additionally require numerical upper bounds
for the score radii and a lower bound for the edge signal.

The first line of the bound retains the intersection between score accuracy
and separation failure and is at least as sharp as its union-bound form in the
second line. Neither bound requires independence among the events. A uniform
signal condition that places \(\lambda_S\) inside the separating interval
makes the final term zero. {For any empirical training-stage selection,
the finite-sample theorem applies directly only when it returns the population
candidate sets and model variables; Proposition~\ref{prop:training-selection-recovery}
instead treats the realized training choices under coverage, accuracy, and
gap conditions.}

The conditional-distribution estimator determines the fitting component of
the recovery radius through an exact risk identity. For one comparison,
suppress the pair and component indices and let \(K_0\) and \(K_1\) be the
expected conditional KL risks of the reduced and augmented fitted
distributions relative to their true counterparts, conditional on training.
Then Supplementary Proposition~S2 gives
\[
\E(\widehat\Delta\mid\mathcal F_{\mathrm{tr}})-\Delta=K_0-K_1,
\qquad
\bigl|\E(\widehat\Delta\mid\mathcal F_{\mathrm{tr}})-\Delta\bigr|
\le K_0+K_1.
\]
The expectations defining these risks are over a fresh observation from the
study population. Thus the bias depends on the difference in fit quality;
using the same fitting procedure in both models promotes comparability while
leaving their risk difference to be controlled. If their summed risk is
uniformly bounded by
\(r^{\mathrm{fit}}_{c,n_{\mathrm{tr}}}(\delta_c/2)\), that quantity may replace
\(\varepsilon_c^{\mathrm{app}}+
\rho_{c,n_{\mathrm{tr}}}(\delta_c/2)\) in \(B_c\). An estimator with a
uniform risk rate \(r_n\) yields score error of order
\(r_n+\sqrt{\log(M_c)/n_{\mathrm{val}}}\), up to the Bernstein remainder,
and exact recovery follows when the minimum active-edge signal dominates twice
this radius. Supplementary Corollary~S4 states the
resulting graph-error probability.

For a nonempty edge set, the same reasoning quantifies the effect of
approximate proxy adjustment.
Replace the exact nonedge nulls by the blocking premise of
Proposition~\ref{prop:latent-imbalance}, and let \(\xi_{\max}\) bound
\(\xi_{jr}^c\) over the eligible nonparents and all included comparisons,
taking \(\xi_{\max}=0\) when there are no eligible nonparents.
On the exact-selection and uniform-accuracy events, every estimated nonparent
gain is at most \(\xi_{\max}+B_S\), while every estimated parent gain is at
least \(\kappa_S^{\mathrm{tr}}-B_S\). Hence
\begin{equation}
\kappa_S^{\mathrm{tr}}>\xi_{\max}+2B_S,
\qquad
\xi_{\max}+B_S<\lambda_S<\kappa_S^{\mathrm{tr}}-B_S
\label{eq:imbalance-risk-gap}
\end{equation}
is sufficient for exact graph recovery, as formalized in Supplementary
Corollary~S10. More accurate density estimation reduces
\(B_S\); better proxy adjustment can reduce \(\xi_{\max}\). Increasing the
sample size addresses estimation error, whereas a positive {residual-dependence
bound} remains a distinct identification limitation. This recovery condition
permits approximate separation; the null-based testing guarantee below
requires its own component-null validity.

{Equation~\eqref{eq:imbalance-risk-gap} also gives an operational
sensitivity analysis. Let \(b\ge0\) be a scientifically prespecified upper
bound on the largest nonparent gain and let \(B_S\) be a justified uniform
score-error radius. Retain an eligible direction when
\(\widehat\Omega_{jr}>b+B_S\), followed by acyclic projection. On the candidate-coverage and uniform score-accuracy events for the
prescribed comparison variables, this rule
recovers the graph whenever \(b\) bounds the nonparent scores and every parent
gain exceeds \(b+2B_S\). Varying \(b\) reports how the graph depends on the
allowed departure from {exact proxy separation}. Because {latent dependence is}
unobserved, \(b\) is a sensitivity input rather than an estimate; without a
numerical error radius, the resulting graph path is descriptive. This
threshold analysis is distinct from the implemented zero-null BH/Holm rules.}

The supplementary theory separates five extensions: lower-confidence-bound
ancestor screening (Proposition~S1),
lower-confidence-bound edge selection and responsive-edge recovery
(Corollary~S6), approximate nonedge separation
(Corollary~S10), {training-stage selection}
(Proposition~\ref{prop:training-selection-recovery}), and a fixed-dimensional
logistic risk illustration (Corollary~S11). The last is
{specific to the stated fixed-dimensional logistic setting}; flexible
estimators require their own risk-rate analysis.

\subsection{Simultaneous Error Control after Training-Stage Selection}

The global Holm {method} uses the three families that are fixed conditional
on \(\mathcal F_{\mathrm{tr}}\),
\(\mathcal H^S_{\mathrm{tr}}\), \(\mathcal H^R_{\mathrm{tr}}\), and
\(\mathcal H^C_{\mathrm{tr}}\) defined in Section~3. Conditional on
\(\mathcal F_{\mathrm{tr}}\), a component p-value is super-uniform when
\begin{equation}
\Pbb(\pi^c_{jr}\le t\mid\mathcal F_{\mathrm{tr}})\le t
\quad\text{for every }t\in[0,1].
\label{eq:conditional-component-validity}
\end{equation}

{For a pair with an active assignment comparison, the fitted comparisons
test the training-selected population nulls
\(H^{X,\mathrm{sel}}_{0,jr}:\Delta^{X,\mathrm{sel}}_{jr}=0\) and
\(H^{U,\mathrm{sel}}_{0,jr}:\Delta^{U,\mathrm{sel}}_{jr}=0\). The structural
and concordance nulls are, respectively,}
\[
{H^{S,\mathrm{sel}}_{0,jr}=H^{X,\mathrm{sel}}_{0,jr}
\cap H^{U,\mathrm{sel}}_{0,jr},
\qquad
H^{C,\mathrm{sel}}_{0,jr}=H^{X,\mathrm{sel}}_{0,jr}
\cup H^{U,\mathrm{sel}}_{0,jr}.}
\]
Therefore \(2\min(\pi^X_{jr},\pi^U_{jr})\), truncated at one, is valid for
the structural null when both component p-values are valid;
\(\pi^U_{jr}\) tests intervention responsiveness; and
\(\max(\pi^X_{jr},\pi^U_{jr})\) is valid for the concordance null when at
least one null component p-value is valid. If the assignment comparison is
inactive, the structural null reduces to
\(H^{X,\mathrm{sel}}_{0,jr}\). {On the exact-selection event,
Assumption~2 maps every graph nonedge to these selected-comparison nulls.
The theorem below uses a training-measurable event to state the corresponding
requirement for general training-stage selection.}

\begin{theorem}[Familywise error control]
\label{thm:dual-fwer}
Suppose the tested families and all objects generating
\(\mathcal F_{\mathrm{tr}}\) are fixed before validation outcomes are used.
Let \(\mathcal G_{\mathrm{cal}}\in\mathcal F_{\mathrm{tr}}\) be an event such
that, for almost every training realization in
\(\mathcal G_{\mathrm{cal}}\), the following conditions hold conditional on
\(\mathcal F_{\mathrm{tr}}\):
\begin{enumerate}[label=(\roman*),leftmargin=2em]
\item for each training-eligible structural nonedge in
\(\mathcal H^S_{\mathrm{tr}}\), \(\pi^X_{jr}\) is super-uniform and, when
\(\widehat g_{\mathrm{tr}}\widehat a^U_{jr}=1\), \(\pi^U_{jr}\) is also
super-uniform;
\item for each tested pair in \(\mathcal H^R_{\mathrm{tr}}\setminus E^R\),
\(\pi^R_{jr}=\pi^U_{jr}\) is super-uniform; and
\item for each tested pair in \(\mathcal H^C_{\mathrm{tr}}\setminus E^C\),
at least one of \(\pi^X_{jr}\) and \(\pi^U_{jr}\) is super-uniform.
\end{enumerate}
Then, for prespecified \(\alpha_S,\alpha_R,\alpha_C\in(0,1)\), and
{almost every} training realization in \(\mathcal G_{\mathrm{cal}}\), Holm adjustment
controls conditional FWER at \(\alpha_S\) for
\(\widehat{\mathcal R}^S\), at \(\alpha_R\) for
\(\widehat{\mathcal R}^R\), and at \(\alpha_C\) for
\(\widehat{\mathcal R}^C\). Acyclic projection preserves structural FWER.
The conditional probability of any false declaration across the three
families is at most
\(\alpha_\Sigma=\min\{1,\alpha_S+\alpha_R+\alpha_C\}\). If, for a
deterministic \(\delta_{\mathrm{cal}}\in[0,1]\),
\(\Pbb(\mathcal G_{\mathrm{cal}}^c)\le\delta_{\mathrm{cal}}\), then structural
FWER is at most
\[
\delta_{\mathrm{cal}}+(1-\delta_{\mathrm{cal}})\alpha_S,
\]
and the probability of any false declaration across the three families is at
most
\[
\delta_{\mathrm{cal}}+(1-\delta_{\mathrm{cal}})
\alpha_\Sigma.
\]
When
\(\widehat g_{\mathrm{tr}}=0\), the responsive and concordance families are
empty.
\end{theorem}

Conditional on \(\mathcal G_{\mathrm{cal}}\), family budgets satisfying
\(\alpha_S+\alpha_R+\alpha_C\le\alpha\) control the probability of any false
declaration at level \(\alpha\). Unconditional level-\(\alpha\) control also
requires
\[
\delta_{\mathrm{cal}}+(1-\delta_{\mathrm{cal}})\alpha_\Sigma\le\alpha.
\]
For \(\delta_{\mathrm{cal}}<1\), a useful sufficient allocation has
\(\delta_{\mathrm{cal}}\le\alpha\) and
\(\alpha_S+\alpha_R+\alpha_C
\le(\alpha-\delta_{\mathrm{cal}})/(1-\delta_{\mathrm{cal}})\).
When the families are reported separately, each unconditional guarantee has
the corresponding calibration-failure allowance.

Holm adjustment does not require independence among the p-values. Acyclic
projection preserves structural FWER because it only removes accepted
directions. When {the population variables \(Z^c_{jr}\) are used},
Assumption~\ref{ass:proxy-sufficient}(a)--(b) supplies the component nulls;
{after training-stage selection, validity is required for the actual fixed
families of comparisons}. The failure probability
\(\delta_{\mathrm{cal}}\) measures how often training produces at least one
comparison for which a graph nonedge lacks the required component null or its
p-value calibration. Such a failure can arise from omitted path-blocking
variables, an incompatible proxy-adjustment set, inadequate assignment support,
or a misspecified calibration model. Sample splitting and Holm adjustment do
not bound this probability. Exact recovery of the population ancestor sets,
together with Assumptions~1--2 and valid component tests, is one sufficient
route to \(\mathcal G_{\mathrm{cal}}\); it is stronger than the theorem
requires. Other selection rules need their own bound on
\(\delta_{\mathrm{cal}}\).

\begin{remark}[Normal-reference validity]
Under a nonedge null, consistent reduced and augmented fits approach the same
conditional distribution, so their log-likelihood-ratio variance may shrink.
Normal-reference validity therefore requires {control after division by the estimated standard error.}
Let \(\mu^c_{jr,n}\) and \((\sigma^c_{jr,n})^2\) denote the conditional
mean and variance of one validation log-likelihood ratio, given training.
Supplementary Corollary~S8 gives
fixed-\(p\) asymptotic validity under a conditional central limit theorem,
relative consistency of the standard-deviation estimator, and
\(\sqrt{n_{\mathrm{val}}}\,(\mu^c_{jr,n})_+/\sigma^c_{jr,n}=o_p(1)\)
uniformly over component nulls, where \(x_+=\max(x,0)\). Only upward bias
requires this control for a one-sided test; downward bias can make the test
conservative. The absolute score-risk bound above is sufficient when it is
negligible relative to \(\sigma^c_{jr,n}/\sqrt{n_{\mathrm{val}}}\).
More sharply, under a component null the relevant bias is
\((K_0-K_1)_+\), the positive difference between the two conditional KL
risks. The supplement gives a Gaussian regression example in which the
reduced distribution is known and the condition holds even as the score
variance vanishes. With two estimated distributions, vanishing KL risks alone
are insufficient: if their positive difference and the score standard deviation
have orders \(n_{\mathrm{tr}}^{-1}\) and \(n_{\mathrm{tr}}^{-1/2}\), respectively,
the standardized bias has order
\(\sqrt{n_{\mathrm{val}}/n_{\mathrm{tr}}}\). A fixed training--validation ratio
therefore requires additional control of this positive risk difference.
Establishing the
one-sided risk condition and the conditional limit remains specific to the
fitting procedure, including the boosting implementation studied here.
\end{remark}

Gaussian linear nulls also admit an exact conditional calibration {of likelihood comparisons whose fitted models are fixed using training
data, with both means estimated from training.}
Supplementary Section~S2.5 derives
this reference distribution; Section~\ref{subsec:gaussian-validation}
examines the resulting graph decisions. {For a training-fixed comparison,
it is the one-sided partial-regression \(t\) test whose direction is selected
by the sign of the training coefficient. It assumes a linear Gaussian null,
{independent validation errors with constant variance}, a full-rank reduced design, and at
least two residual degrees of freedom; the reference is conditional on the
validation design and nuisance sufficient statistics. The fitted coefficient
magnitude and the reduced fitted coefficients cancel from this conditional
reference. This exact calibration differs from applying a normal tail directly
to the mean held-out likelihood gain.}

\section{Simulation Studies and K562 Perturb-seq Analysis}

\noindent\textbf{Scope of numerical evidence.}
The paired continuous diagnostic in Section~\ref{subsec:adaptive-comparison}
evaluates the two fixed component/calibration settings together with PABM's
training-selected setting. The
displayed experiments assess full-DAG
recovery in their stated designs. Section~\ref{subsec:gaussian-validation}
additionally evaluates structural false-edge control for a Gaussian
implementation. The responsive and concordance Holm families are computed in the
continuous implementation, but their recovery and error-control performance
are not assessed in the reported experiments.

We evaluate full-DAG recovery using simulations with unmeasured
confounding and illustrate the procedure in K562 CRISPRi Perturb-seq data.
The simulation evidence consists of low-dimensional continuous models and a
measurement-layer count benchmark with structural log-rate DAGs observed
through Poisson counts.
When a reference DAG is known, our primary measure is structural Hamming
distance (SHD), which counts missing, extra, and reversed directed edges in
the reconstructed graph. {A reversal costs one SHD operation. In directed
precision and recall, the incorrectly oriented arrow is a false positive and
the absent correct orientation is a false negative.} We also report directed F1, the harmonic mean of
directed-edge precision and recall, as a secondary selected-graph diagnostic.
For the count study, average precision (AP) evaluates the {directed-edge ranking before edges are selected. Average precision
summarizes precision across recall levels; nonfinite or unevaluated scores
are assigned the same score below all finite scores.} Thus SHD is the principal recovery
criterion, directed F1 corroborates selected directed-edge retrieval, and AP
assesses ranking before graph selection.

\subsection{{Simulations with Gaussian Responses: Selection and Error Control}}
\label{subsec:gaussian-validation}

We first examine a fully specified Gaussian implementation of PABM and the
adaptive implementation used in the subsequent benchmarks. The six-node DAG
has the paths \(X_1\to X_2\to X_3\to X_6\) and
\(X_1\to X_4\to X_5\to X_6\). Generate
\(X_r=0.8\sum_{j\in\Pa r}X_j+U_r+0.7H+\varepsilon_r\), with mutually
{independent standard normal assignments, confounder, and node-specific
noise terms.}
The exact-adjustment reference records \(W=H\); the matched noisy-proxy setting
records \(W=H+0.75\zeta\), with an independent standard normal \(\zeta\).
Each uses 400 paired replicates with 700 training and 300 validation observations.
Assumptions~1--2 hold for the population ancestor comparisons in the
exact-adjustment setting. The noisy setting examines departure from nonedge
separation.

The Gaussian implementation uses training-only lower confidence bounds for
ancestor gains, least-squares conditional means, both direct-edge comparisons,
and global Holm adjustment at level \(0.05\). We compare conditional Gaussian
calibration of the {held-out gains from training-fixed models} with their working normal-reference
calibration, keeping the candidates and fits identical.
Supplementary Section~S2.5 derives
the confidence bounds and the exact conditional reference distribution.
The adaptive boosting implementation uses the candidate, tuning,
component-selection, and BH/Holm rules. All three receive the same observations;
truth is used only after fitting for evaluation.

\begin{table}[!tbp]
\centering
\small
\setlength{\tabcolsep}{4pt}
\begin{tabular}{llccc}
\toprule
Proxy & Fit and calibration & SHD (SE) & Any false edge [95\% CI] & {Exact final ancestor sets} \\
\midrule
Exact & Gaussian, conditional & 0.050 (0.011) & 0.050 [0.033, 0.076] & 400/400 \\
 & Gaussian, normal & 0.033 (0.009) & 0.033 [0.019, 0.055] & 400/400 \\
 & Adaptive boosting & 0.843 (0.035) & 0.693 [0.646, 0.736] & 0/400 \\
\addlinespace
Noisy & Gaussian, conditional & 0.802 (0.045) & 0.542 [0.494, 0.591] & 390/400 \\
 & Gaussian, normal & 0.235 (0.023) & 0.215 [0.178, 0.258] & 390/400 \\
 & Adaptive boosting & 0.985 (0.037) & 0.767 [0.724, 0.806] & 0/400 \\
\bottomrule
\end{tabular}
\caption{Gaussian validation on a six-edge DAG. SHD entries are means with
Monte Carlo standard errors conditional on the fixed graph and coefficients.
{Here SE denotes Monte Carlo standard error and CI denotes confidence
interval. The false-edge column is the frequency} of at
least one false edge after acyclic projection, with Wilson 95\% intervals;
the last column counts repetitions {whose final candidate sets equal all
population ancestor sets}.
Every candidate set contained all true parents in every repetition. Gaussian rows use the same
LCB selection and global Holm; adaptive boosting uses its prespecified empirical
selection and adjustment rules.}
\label{tab:gaussian-validation}
\end{table}

With exact adjustment, the Gaussian implementation recovered all ancestor
sets in every repetition. Conditional calibration gave a false-edge frequency
of \(0.050\), with interval \([0.033,0.076]\), and mean SHD \(0.050\).
The matched normal-reference rule had a lower observed false-edge frequency.
{For this specified design, the analytic union bound in Supplementary
Section~S2.5 is
\(\delta_{\rm anc}\le 0.00793\): its extra-ancestor term is
\(19/2400=0.00792\), and the sum of the 11 noncentral-\(t\) ancestor-miss
probabilities is \(4.38\times10^{-6}\). The resulting unconditional
structural-FWER bound is \(0.05+0.95\delta_{\rm anc}\le0.0576\). The observed
400/400 selections are consistent with this calculation but do not establish
selection probability one.}
The adaptive implementation had mean SHD \(0.843\) and a false-edge frequency
of \(0.693\). It used the expression-only, targetwise-BH route throughout,
{used the replacement candidate rule in 298 of 400 repetitions,
which selects the top-ranked candidates when the initial screen retains
too few on average,} and never recovered all
ancestor sets. {This zero frequency is forced by the fallback policy rather
than evidence that the initial residual screen always failed. The six targets
have \(0,1,2,1,2,5\) strict ancestors, totaling 11. An exact initial screen
therefore averages \(11/6<2\) candidates and activates the fallback, which
uses \(\min(20,p-1)=5\) candidates per target and returns all 30 ordered
pairs. If the initial screen instead averages at least two, retaining it gives
at least 12 pairs. Under either branch, the final family cannot equal the
11-pair ancestor family.} Although its candidate sets contained every parent, its selected regression
covariates left a mean largest nonparent population gain of \(0.117\).
{Its mean targetwise false-discovery proportion, averaging equally over
targets and repetitions, was 0.066 (Monte Carlo SE 0.004); this is the
error criterion corresponding most closely to its targetwise BH rule and
exceeds the nominal 0.05 level. The fallback admits descendants among the
regression covariates; conditioning on a common descendant of modeled responses
can open a collider path even after exact adjustment for \(H\). For example,
conditioning on \(X_6\) in the subgraph \(X_3\to X_6\leftarrow X_5\) can
associate the nonadjacent variables \(X_3\) and \(X_5\). Such induced
associations yield positive nonparent gains and invalidate the BH null
p-values. The
69.3\% graph-level false-edge frequency is a separate summary and is not a
test of a targetwise false discovery rate (FDR) claim.} Thus parent coverage
alone can leave graph
nonedges conditionally associated.
These configurations differ in selection, fitting, and adjustment; the
comparison does not isolate the effect of boosting.

With noisy proxies, the Gaussian selector recovered all ancestor sets in
390 repetitions, yet conditional calibration gave a false-edge frequency
of \(210/390=0.538\) among those repetitions. Residual dependence violates
the graph nonedge nulls even with the correct ancestor sets. This diagnostic
connects a concrete Gaussian implementation to the calibration guarantee and
separates its requirements from empirical graph-recovery performance.
{The adaptive route's mean targetwise false-discovery proportion was
0.077 (Monte Carlo SE 0.004).}

\subsection{Full-DAG Recovery: Protocol and Results}
\label{subsec:uniform-comparison}

\paragraph{Common protocol.}
{Every method is evaluated on the same seed-specific replicate and sample
budget, using the inputs supported by its documented interface. The continuous panels use a common 70\%/30\%
training--validation split. In the K562-calibrated count panels, PABM uses its
{prespecified guide-stratified 70\%/30\% roles, splitting cells separately
within each guide-assignment group:} training data construct and fit
the candidate-specific comparisons, and validation data evaluate the direct
\CLLR{} statistics. {ARGEN and the released GAMPI implementation used here (GAMPI-DRI) receive all rows} through their
documented released interfaces; NOTEARS uses the {same prespecified split} for its
validation-based graph tuning. {When a PABM conditional {estimator} has
tuning parameters, selection uses training observations only. The continuous
studies use three-fold cross-validation; the count study uses a deterministic
80\%/20\% split within the training sample. Both select one direct-edge
configuration per replicate, which is then refitted on all training
observations.} {Each competitor retains its documented model class and is
evaluated through the {method-specific pipeline specified in the supplement, with fitting and
graph selection performed without using the true graph or its edge count.} ARGEN and GAMPI-DRI use released interfaces; DeFuSE, NOCADILAC,
and NOTEARS use {documented validation or threshold-selection procedures needed}
to produce automatic DAG estimates. The comparison therefore concerns these
specified pipelines rather than unmodified native selection rules.} We use the authors' released code
when available and {document the implementation changes needed to run the released code and
the automatic-DAG rules} in Supplementary
Section~S1.1.2. No procedure uses the true
graph or true edge count during fitting or automatic graph selection.}

{In the following benchmark panels, for every training-selected candidate,
PABM fits the balanced
comparisons of Section~3 using histogram gradient boosting with a Gaussian or
Poisson working distribution and evaluates their \CLLR{} scores on validation data
\citep{friedman2001gradient,ke2017lightgbm}. The reported candidate selector
uses the {empirical residual-association screen}. When that screen averages fewer than
two candidates per target, we use the {empirical top-\(\min(20,p-1)\)
fallback candidate selector}, which replaces each target's residual-screen set
by its {top-\(\min(20,p-1)\)} training-only
nonlinear-association set. The expression-only branch uses targetwise BH; the dual branch uses
Bonferroni-minimum component p-values with global Holm; both use greedy
acyclic projection. These are empirical implementation choices distinct from
the lower-confidence-bound screen and formal calibration assumptions.
Supplementary Section~S1.2 gives the
transformations, thresholds, included variables, fitting grids, and tie
rules. No screening or graph-recovery theorem is asserted for this
fallback variant.}

The comparison includes NOTEARS as an observed-variable linear-DAG baseline
\citep{zheng2018notears}. {We use GAMPI-DRI as the representative IV-based
comparator because it extends the GrIVET framework to generalized-linear
outcomes and adjusts for unmeasured confounding
\citep{chen2024grivet,wang2024gampi}.} DeFuSE represents nonlinear discovery
with correlated Gaussian confounding \citep{li2024nonlinear}, and NOCADILAC
represents nonlinear unmeasured-confounder modeling
\citep{kaltenpoth2023nocadilac}. ARGEN is included in the count study, where its
perturbation inputs are appropriate
\citep{park2026argen}. {The
simulation tables report Monte Carlo means with standard errors \(s/\sqrt R\)
in parentheses. Paired simulation comparisons use the same seeds and report
paired 95\% \(t\)-intervals.}

Throughout the following benchmarks, \(H_i\) denotes the unobserved
shared confounder block, \(U_i\) the recorded intervention-assignment block,
and \(W_i\) the recorded covariate and proxy block. Proxy measurements are
components of \(W_i\), never components of \(H_i\). Thus PABM receives
\((X_i,U_i,W_i)\), whereas \(H_i\) appears only in the data-generating
mechanism.

\paragraph{Continuous designs.}
The low-dimensional comparison uses \(p=50\), \(n=1000\), and 20 paired
seeds. Table~\ref{tab:uniform-comparison} reports the training-selected PABM
setting; {Supplementary Table~S8} compares the two fixed ablations
on separate paired cases.
\begin{samepage}
All three continuous panels follow
\[
X_{ir}=g_r\bigl(X_{i,\Pa{r}}\bigr)
       +0.65H_i+0.60U_{ir}+\varepsilon_{ir},
\]
The linear and cosine panels share a generated DAG and its structural
coefficients and differ in the causal function \(g_r\). The nonadditive panel
uses a separate DAG from the same ordered maximum-indegree-two generator:
\begin{align}
g_r^{\mathrm{lin}}(X_{i,\Pa{r}})
&=\sum_{j\in\Pa{r}}{A}_{jr}X_{ij},
&&\text{linear},\label{eq:simulation-linear}\\
g_r^{\mathrm{cos}}(X_{i,\Pa{r}})
&=2\sum_{j\in\Pa{r}}{A}_{jr}\cos(X_{ij}),
&&\text{nonlinear additive},\label{eq:simulation-nonlinear}\\
g_r^{\mathrm{int}}(X_{i,\Pa{r}})
&=1.5\left(
  \sum_{j\in\Pa{r}}{A}_{jr}\tanh(X_{ij})
  +\gamma_r\!\prod_{\ell=1}^{2}\tanh(X_{i j_{r\ell}})
  \right),
&&\text{nonlinear nonadditive},
\label{eq:simulation-nonadditive}
\end{align}
\end{samepage}
{Conditional on \(X_{i,\Pa{r}},H_i,U_{ir}\), this is a Gaussian
{additive-noise structural equation model (SEM):}
\[
X_{ir}\mid X_{i,\Pa{r}},H_i,U_{ir}
\sim N\!\left(g_r(X_{i,\Pa{r}})+0.65H_i+0.60U_{ir},\,0.55^2\right).
\]
Thus the first panel is a linear-Gaussian SEM, whereas the cosine and
interaction panels are nonlinear Gaussian additive-noise SEMs whose observed
joint distributions may be non-Gaussian. Here,}
\(H_i,U_{ir}\sim N(0,1)\),
\(\varepsilon_{ir}\sim N(0,0.55^2)\), and the nonzero \({A}_{jr}\) have
random signs and magnitudes drawn uniformly between 0.80 and 0.95. In the
third mechanism, \(j_{r1}\) and \(j_{r2}\) are the two parents of node \(r\),
and every two-parent node has a signed interaction coefficient with
\(|\gamma_r|\) drawn uniformly between 0.60 and 0.80; \(\gamma_r=0\) otherwise.
The multiplicative term makes the contribution of either parent depend on the
value of the other and therefore violates additivity across parents.
{The recorded auxiliary block is \(W_i=(W_{i1},W_{i2})\), with proxy coordinates
\(W_{i1}=H_i+0.35\eta_{i1}\) and
\(W_{i2}=0.70H_i+0.35\eta_{i2}\), where
}
\(\eta_{i1},\eta_{i2}\overset{\mathrm{iid}}{\sim}N(0,1)\) independently of
\(H_i\), \(U_i\), and \(\varepsilon_i\). {Within each replicate, the
means and standard deviations of the nodes, intervention assignments, and
proxy measurements are computed from the training observations; the same
training values are used to standardize validation observations.}
{PABM uses \(X,U,W\). GAMPI-DRI receives \(X,U\) and treats \(U\) as
IVs; its released interface does not accept \(W\). DeFuSE, NOCADILAC, and
NOTEARS receive the observed nodes \(X\).}
{The same histogram-boosting {estimator class} and tuning grid are
prespecified for PABM throughout these three continuous panels. Within each
replicate, cross-validation selects one direct-edge
configuration from prespecified representative targets and refits it for
every evaluated target--candidate pair.}

{Supplementary Figure~S1 displays the
97-edge DAG shared by the linear and cosine panels. The nonadditive panel
uses a separately drawn DAG from the same generator. All three panels use
the same sample size, split ratio, and five methods.}

\begin{table}[!tbp]
\centering
\begingroup
\setlength{\tabcolsep}{7pt}
\small
\begin{tabular}{llcc}
\toprule
Structural mechanism & Method & Mean SHD (SE) $\downarrow$ & Mean F1 (SE) $\uparrow$ \\
\midrule
Linear Gaussian
& {PABM} & \textbf{24.55 (0.93)} & \textbf{0.868 (0.005)} \\
& GAMPI-DRI & 121.00 (4.53) & 0.527 (0.016) \\
& DeFuSE & 97.00 (0.00) & 0.000 (0.000) \\
& NOCADILAC & 274.60 (12.35) & 0.039 (0.005) \\
& NOTEARS & 182.40 (4.76) & 0.353 (0.008) \\
\addlinespace
Nonlinear cosine
& {\makecell[l]{PABM + empirical\\top-20 fallback}} & \textbf{26.40 (0.92)} & \textbf{0.856 (0.005)} \\
& GAMPI-DRI & 96.40 (1.26) & 0.230 (0.011) \\
& DeFuSE & 87.85 (0.88) & 0.170 (0.015) \\
& NOCADILAC & 272.45 (12.98) & 0.040 (0.005) \\
& NOTEARS & 284.90 (6.83) & 0.108 (0.005) \\
\addlinespace
Nonadditive interaction
& {PABM} & \textbf{9.00 (0.48)} & \textbf{0.952 (0.003)} \\
& GAMPI-DRI & 65.30 (3.32) & 0.733 (0.012) \\
& DeFuSE & 97.00 (0.00) & 0.000 (0.000) \\
& NOCADILAC & 252.10 (4.25) & 0.056 (0.004) \\
& NOTEARS & 194.15 (1.96) & 0.361 (0.004) \\
\bottomrule
\end{tabular}
\endgroup
\caption{PABM low-dimensional graph-recovery comparison for five methods with graph size \(p=50\),
sample size \(n=1000\), a common 70\%/30\% training--validation split, and 20
paired simulation replicates for each row. The PABM rows use the training-only
{{empirical residual-association screen and component-and-calibration-route rule}; the cosine
panel uses the explicitly labeled PABM variant with the {empirical
top-20 fallback candidate selector}. The rows then use} held-out direct
\CLLR{} scoring; comparator rows use
{their prespecified fitting and graph-selection procedures, which do not
use the true graph.} Each method uses
{its prespecified automatic graph-selection rule without access to the
true graph.} Values are Monte Carlo means with standard errors in parentheses. SHD
denotes structural Hamming distance, and F1 is the harmonic mean of
directed-edge precision and recall. Boldface marks the best mean within each
structural mechanism. {The Monte Carlo variation is conditional on the
fixed graph and coefficients used for each mechanism.}}
\label{tab:uniform-comparison}
\end{table}

{Table~\ref{tab:uniform-comparison} evaluates automatically selected
DAGs at a common data budget. {PABM is used in
the linear and nonadditive panels and PABM with the {empirical top-20
fallback candidate selector} in the cosine panel; these rows have the lowest mean
SHD and the highest mean directed F1 in their respective panels.}
It retains both comparisons in all 20 linear and all 20 nonadditive fits;
all 20 cosine fits use only the expression comparison. The {component-and-calibration-route rule} is computed on training
data exclusively, leaving the held-out direct \CLLR{} scores for edge
evaluation. {{Supplementary Table~S8} reports a separate
configuration comparison.}}

{The DeFuSE selection adapter returns an empty graph for every linear and
nonadditive replicate: all retained threshold candidates are empty, although
the archive does not establish that the unthresholded network weights vanish.
The NOCADILAC adapter returns its final iterate because its released code has
no compatible best-iterate checkpoint. These rows therefore describe the
specified pipelines rather than intrinsic failure of the underlying model
classes; the supplement gives the relevant implementation details.}

To assess recorded proxy information, we rerun PABM after permuting \(W\)
within each sample split and after omitting \(W\). Cases, splits, tuning, and
seeds remain fixed; each diagnostic recomputes every data-dependent choice.

\begin{table}[!tbp]
\centering
\begingroup
\setlength{\tabcolsep}{5pt}
\footnotesize
\begin{tabular}{llcc}
\toprule
Structural mechanism & Diagnostic minus full \(W\) & \(\Delta\)SHD & \(\Delta\)F1 \\
\midrule
Linear Gaussian
& Permuted \(W\) & 20.70 [17.99, 23.41] & $-.113$ [$-.127$, $-.099$] \\
& Omitted \(W\) & 19.15 [16.13, 22.17] & $-.105$ [$-.121$, $-.090$] \\
Nonlinear cosine
& Permuted \(W\) & 6.90 [3.01, 10.79] & $-.032$ [$-.050$, $-.013$] \\
& Omitted \(W\) & 8.00 [4.46, 11.54] & $-.036$ [$-.052$, $-.020$] \\
Nonadditive interaction
& Permuted \(W\) & 13.45 [11.10, 15.80] & $-.073$ [$-.086$, $-.059$] \\
& Omitted \(W\) & 13.95 [11.41, 16.49] & $-.076$ [$-.090$, $-.062$] \\
\bottomrule
\end{tabular}
\endgroup
\caption{Paired sensitivity to recorded proxy information over {20 prespecified}
replicates per mechanism. Entries are mean diagnostic-minus-full-\(W\)
differences with paired 95\% intervals. Positive \(\Delta\)SHD and negative
\(\Delta\)F1 favor full \(W\). The full-\(W\) arm is PABM; permuted and omitted
\(W\) are complete-procedure diagnostics.
Monte Carlo variation is conditional on the fixed graph and coefficients.
Supplementary Table~S1 reports absolute metrics, standard
errors, false positives, recall, precision, and selected graph sizes.}
\label{tab:proxy-quality-ablation-main}
\end{table}

Table~\ref{tab:proxy-quality-ablation-main} shows that both diagnostics raise
SHD and reduce directed F1 in every mechanism; all six SHD and six F1 intervals
exclude zero. In the cosine mechanism, recall is nearly unchanged and the SHD increase
is driven primarily by additional false positives. Proxy degradation reduces
both recall and precision in the linear and nonadditive mechanisms. Because
\(W\) enters the ancestor screen and parent comparisons, the experiment
measures complete-procedure sensitivity to recorded proxy information. It
neither isolates one stage, estimates the {latent dependence in}
Proposition~\ref{prop:latent-imbalance}, nor verifies Assumption~2.

{
\subsection{Training-Based Selection of Comparisons and Calibration Rules}
\label{subsec:adaptive-comparison}

{This diagnostic compares the training-selected component and calibration
route with two fixed ablations. The expression route uses targetwise BH and
greedy projection; the dual route uses Bonferroni-minimum p-values, global
Holm, and p-value-ranked projection. Each panel has 20 cases with \(p=50\),
\(n=1000\), and a 70\%/30\% split. The empirical training rule and all
preprocessing details appear in Supplementary
Section~S1.2.}

{The training rule selects the lower-SHD route in all three designs:
dual/global-Holm in the linear and nonadditive panels, and
expression/targetwise-BH in the cosine panel. Supplementary
Table~S8 gives the SHD, F1, and paired changes.
Holding global Holm fixed in the paired linear and cosine cases likewise
shows that adding assignment evidence improves the former and worsens the
latter. Its incremental value is therefore design-dependent and retains the
stated exclusion, support, and nonedge-validity requirements.}
}

\subsection{\texorpdfstring{{Simulations with Count Responses Calibrated to K562}}{Simulations with Count Responses Calibrated to K562}}
\label{subsec:count-main}

{These simulations retain the observed study design and covariates while
generating a known graph of unobserved expression rates and new counts.
The retained inputs come from the released ARGEN chromosome-6 K562
Perturb-seq panel}
\citep{park2026argen}. It retains all 19,251 released cell rows, 45 selected
genes, 45 one-hot guide assignments, 10,691 non-targeting controls, the
released exposure offset, the 47-column gel bead-in-emulsion (GEM)
auxiliary-measurement block, and
the mitochondrial-percentage technical covariate. The observed target-arm
sizes range from 101 to 485 cells. Thus the benchmark retains the cell count,
guide imbalance, depth variation, and recorded covariate and proxy
information of this Perturb-seq design, while generating a known structural
graph independently of the real-data analysis. Its reference DAG is defined
over latent log rates \(R\), whereas Theorem~\ref{thm:population-identification}
targets a graph over the observed variables \(X\). The results therefore
measure recovery of the latent-rate graph through a noisy count layer, a
target distinct from the observed-variable graph in the identification theorem.
{Accordingly, every SHD, F1, and AP value in this subsection is computed
against the latent-rate reference DAG. These metrics quantify recovery through
the Poisson observation layer; no equivalence with the observed-variable
target of Theorem~\ref{thm:population-identification} is asserted.}

{For cell \(i\) and gene \(r\), let \(R_{ir}\) be a structural log expression
rate and let \(X_{ir}\) be the observed count. The fixed truth is a 45-node,
65-edge DAG of the rates \(R_{ir}\), with maximum indegree two. Given the
released exposure \(L_i\), the observation distribution is}
\begin{equation}
\begin{aligned}
X_{ir}\mid(R_{ir},L_i)&\sim
\operatorname{Poisson}\!\left\{L_i\exp(\overline R_{ir})\right\},
\qquad {\overline R_{ir}=\min\{5.8,\max(-7,R_{ir})\},}\\
R_{ir}&=\alpha_r+\sum_{j\in\Pa{r}}b_{jr}g_{jr}(R_{ij}-\alpha_j)
+\lambda_r^{\mathsf T}H_i+\tau U_{ir}
+\psi_r\widetilde W_i^{\mathrm{tech}}+\varepsilon_{ir}+h_r(R_{i,\Pa r}).
\end{aligned}
\label{eq:count-sem}
\end{equation}

{The count comparisons condition explicitly on \(L_i\): their reduced
and augmented fitted distributions are
\(q_r(\,\cdot\mid Z^c_{ijr},L_i)\) and
\(q_r(\,\cdot\mid Z^c_{ijr},V^c_{ij},L_i)\). In this semi-synthetic study {the observed exposure vector \(L\) is held fixed across simulated
replicates.} In the
K562 analysis it is an additional observed cell-level quantity. The general
identification argument extends by appending \(L\) to each displayed
regression-covariate block, provided assignment exogeneity, nonedge
separation, and support continue to hold after conditioning on it. Because
\(L\) is retained, \(X\mapsto\log(1+X/L)\) is invertible for each \(L>0\);
normalizing by an unrecorded random exposure would change the statistical
target.}

{Here \(\alpha_r\) is the fixed control log-rate intercept, \(H_i\) is a
three-dimensional unmeasured confounder, \(\lambda_r\) and \(\psi_r\) are fixed
confounder-loading and technical-covariate coefficients, \(\widetilde
W_i^{\mathrm{tech}}\) is the standardized technical covariate, and
\(\varepsilon_{ir}\) is fresh structural noise. The assigned-intervention
indicator \(U_{ir}\) has fixed targeted-gene effect \(\tau=-0.85\). The parent
log-rate coefficients have magnitudes from 0.45 to 0.65. In the log-linear
panel, \(g_{jr}\) is the identity. In the nonlinear panel, \(g_{jr}\) is a
prespecified bounded \(\tanh\) or sine transformation, and two-parent rates
also include the bounded interaction
\[
h_r(R_{i,\Pa r})=\gamma_r\tanh(R_{ij_1}-\alpha_{j_1})
\sin\{1.5(R_{ij_2}-\alpha_{j_2})\},\qquad \Pa r=\{j_1,j_2\},\ j_1<j_2.
\]
Here \(h_r=0\) for all log-linear rates and for nonlinear rates with fewer
than two parents; \(\gamma_r\) is fixed before fitting as specified in
Supplementary Section~S1.3.
The topology, signs, transformations, and
hidden loadings are fixed before fitting and unavailable to every method.}

The latent-confounder distribution is calibrated from control-cell log
rates. Its three leading empirical components retain their observed
association with released GEM groups; fresh latent confounders, structural
noise, and Poisson draws are generated for each replicate. The complete
recorded covariate and proxy block is
\(W_i=(W_i^{\mathrm{GEM}},W_i^{\mathrm{tech}})\). The GEM measurements serve
as proxy measurements for unmeasured confounding to the extent that they track
the latent confounders,
whereas the technical coordinate enters the rate mechanism directly. The
benchmark therefore evaluates recovery of a structural log-rate DAG observed
through Poisson counts. The unknown K562 gene network lies outside this
truth-known benchmark. The log-linear panel is the closest comparison to ARGEN's
stated structural model, whereas the nonlinear panel assesses workflow
robustness after departure from that log-linear specification.

{Each panel uses 20 matched final seeds. {Within each guide stratum,
a generator initialized at replicate seed plus 8,000,003 permutes the rows;
the first \(\lfloor0.70n_h\rfloor\) enter training and the remainder enter
validation.} This yields 13,451 training and 5,800 validation cells, with 70--339 training and
31--146 validation cells per targeted gene. PABM receives \((X,U,W)\) and the
exposure offset. It forms candidate sets on training cells using the
{empirical residual-association screen}; if that screen averages fewer
than two candidates per target, it substitutes the prespecified
{empirical top-20 fallback candidate selector}. It then computes Poisson {reduced--augmented \CLLR{} scores} on
validation cells. ARGEN receives counts, guide assignments, exposure,
and the same recorded covariate matrix denoted by \(W\) in this paper through
its released covariate interface. PABM treats eligible coordinates as
recorded proxy measurements; ARGEN then constructs its own first-stage
expression proxies from the assignments
and observed covariates; these generated regressors are distinct from the
recorded proxy measurements in \(W\). GAMPI-DRI
receives counts and guide assignments through its released Poisson interface,
and NOTEARS is fitted to {\(\log(1+\text{count}/\text{exposure})\),
standardized using training means and standard deviations}. Their documented interfaces omit
\(W\). {The training-only {component-and-calibration-route rule selects
the expression-plus-assignment/global-Holm route} in all 20 final replicates in
each panel. The empirical residual-association screen retains 141.55 and
107.95 directed candidates on average in the log-linear and nonlinear panels,
respectively.} {No graph-selection rule uses the true graph or its edge count.}}

\begin{table}[!tbp]
\centering
\begingroup
\setlength{\tabcolsep}{3.5pt}
\small
\begin{tabular}{@{}llccc@{}}
\toprule
Structural mechanism & Method & \makecell{Mean SHD (SE)\\$\downarrow$} &
\makecell{Mean F1 (SE)\\$\uparrow$} & \makecell{Mean AP (SE)\\$\uparrow$} \\
\midrule
Log-linear
& {PABM} & 70.75 (1.15) & \textbf{0.473 (0.006)} & 0.416 (0.006) \\
& ARGEN & 365.95 (38.38) & 0.281 (0.019) & \textbf{0.941 (0.006)} \\
& GAMPI-DRI & 234.95 (5.52) & 0.153 (0.004) & 0.080 (0.004) \\
& NOTEARS & \textbf{67.45 (0.17)} & 0.001 (0.001) & 0.081 (0.001) \\
& {Empty-graph reference} & {65.00} & {0.000} & {--} \\
\addlinespace
Nonlinear
& {PABM} & 68.50 (1.16) & \textbf{0.445 (0.006)} & 0.358 (0.008) \\
& ARGEN & 241.65 (21.73) & 0.334 (0.017) & \textbf{0.822 (0.007)} \\
& GAMPI-DRI & 237.90 (6.54) & 0.129 (0.003) & 0.070 (0.002) \\
& NOTEARS & \textbf{68.05 (0.05)} & 0.000 (0.000) & 0.095 (0.001) \\
& {Empty-graph reference} & {65.00} & {0.000} & {--} \\
\bottomrule
\end{tabular}
\endgroup
\caption{{K562-calibrated measurement-layer recovery of the latent-rate
reference DAG through observed Poisson counts for four methods.}
Each panel has 45 structural log-rate nodes, 65 true arrows, 19,251 cells,
and 20 matched final seeds. PABM uses a guide-stratified training--validation
split; ARGEN and GAMPI-DRI use their released all-row interfaces, and NOTEARS
uses normalized counts with split-based tuning. Values are Monte Carlo means
with standard errors in parentheses. SHD denotes structural Hamming distance,
F1 is the harmonic mean of directed-edge precision and recall, and AP is
average precision of the directed-edge ranking. Boldface marks the best mean
{among fitted procedures} within each structural mechanism. {The
empty graph is an evaluation reference rather than a fitted procedure. Monte
Carlo uncertainty is conditional on the fixed latent-rate graph and
coefficients.}}
\label{tab:count-full-dag}
\end{table}

\begin{figure}[!tbp]
\centering
\includegraphics[width=\textwidth]{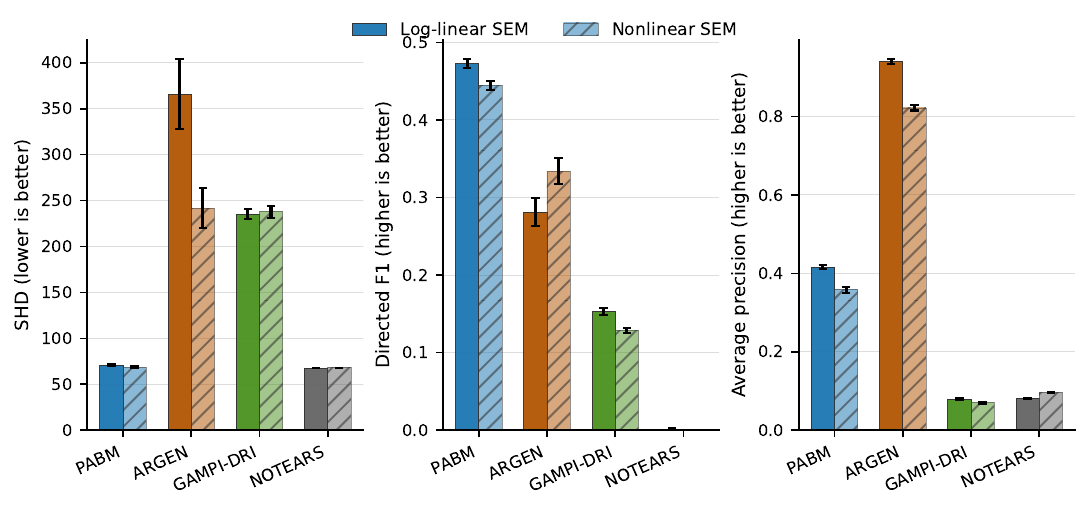}
\caption{K562-calibrated measurement-layer graph-recovery results. Bars are Monte Carlo
means and error bars are standard errors over 20 matched seeds, conditional
on the fixed latent-rate graph and coefficients. {PABM has the}
largest directed F1 among the four displayed methods under both structural
mechanisms. ARGEN has the largest
AP. NOTEARS has the smallest SHD, close to the empty-graph value of 65, and
its automatic graphs contain very few arrows; the accompanying directed F1
panel shows their negligible directed-edge retrieval.}
\label{fig:k562-calibrated-count}
\end{figure}

{Figure~\ref{fig:k562-calibrated-count} and
Table~\ref{tab:count-full-dag} summarize the 40-case comparison.
Relative to the fixed expression/BH ablation, PABM's selected
{expression-plus-assignment setting with global Holm} lowers mean SHD by 8.05 (0.92)
in the log-linear panel and 4.40 (0.97) in the nonlinear panel, and raises AP
by 0.017 (0.001) and 0.010 (0.001), respectively. Its directed F1 is lower by
0.020 (0.007) and 0.024 (0.008). Because both the comparison set and the
multiple-testing rule change, these differences evaluate complete numerical
settings and conflate the component-set and calibration changes. In these K562-calibrated
panels, PABM has much lower SHD than ARGEN and the largest directed F1, while
ARGEN has the largest AP. NOTEARS selects only 2.65 and 3.10 arrows on average;
its mean SHD of 67.45 and 68.05 lies near the empty-graph value of 65, and its
directed F1 is essentially zero. All four methods have mean SHD above the
empty-graph reference of 65. PABM's more selective graphs have higher precision
and directed F1 than ARGEN's, whereas ARGEN recovers more true arrows and has
substantially better {ranking before edge selection}. These results separate automatic
edge selection from ranking and show why SHD, directed-edge retrieval, and AP
are informative together.}

{Evaluation-only diagnostics separate ranking from automatic selection.
At the true graph size of 65, ARGEN has substantially lower SHD than PABM;
PABM's training candidates retain 88.8\% and 72.4\% of the true edges in the
two panels. NOTEARS' selected size is highly sensitive to the dimension scaling
of its edge penalty. Supplementary Table~S7
reports the {curves at common selected graph sizes} and penalty sensitivity analysis. These diagnostics support a
selection--ranking tradeoff rather than a uniform ordering of PABM and ARGEN.}

\subsection{K562 CRISPRi Perturb-seq Illustration}
\label{subsec:k562-main}

We use ARGEN's released chromosome-6 panel of 19,251 K562 cells and
45 genes \citep{park2026argen}. One row is one cell: \(X_i\) is its 45-gene
post-perturbation raw-count vector, and \(U_i\) is the mutually exclusive
45-guide assignment vector, with 10,691 non-targeting cells as the reference
state. The released exposure is the normalized guide unique molecular
identifier (UMI) count,
\(L_i=\texttt{guide\_UMI\_sum}_i/\overline{\texttt{guide\_UMI\_sum}}\) and is
kept separate from \(W_i\). {We preserve this ARGEN-release definition to
align the Poisson offset across the two workflows. It is a guide-capture
quantity rather than the total RNA UMI count, so treating it as a proportional
gene-count exposure is a working measurement assumption.} The 48-coordinate \(W_i\) contains a 47-column
one-hot encoding of the released GEM-group label and the mitochondrial-read
percentage. GEM group is a recorded partition label; mitochondrial percentage
is a contemporaneous quality-control quantity derived from the same cell-level
assay. Both variables lie outside the 45 graph-node counts. Targeted arm sizes
range from 101 to 485 cells.

{These covariates may capture experimental-group and assay-related variation,
but their ability to remove the unmeasured biological dependence relevant
to each candidate edge is not established. The 47 indicator columns encode
one group label, so their number does not establish proxy adequacy. The
contemporaneous mitochondrial measurement also requires justification as
an adjustment variable. Measurements outside the graph-node panel can
still respond to intervention. We therefore use this analysis to illustrate
candidate-network construction and descriptive assessment; it does not
validate the proxy-separation assumptions or establish the selected edges
as direct regulatory relationships.}

{The analysis has no reference DAG. PABM uses a guide-stratified
70\%/30\% split, its empirical residual-association candidate rule, both
supported comparisons, and global Holm; the supplement gives the complete
regression covariates and thresholds. ARGEN uses its released all-cell Poisson
workflow and generated expression proxy. Both methods receive the counts,
guides, exposure, and mitochondrial percentage; PABM additionally receives
47 GEM indicators. In the synthetic count benchmark ARGEN instead receives
the generated complete \(W\) block, so the real-data sensitivity aligns inputs
by removing GEM indicators from PABM. An ARGEN refit adding them was not
performed. Method-specific sample roles and selection remain different.
Cell-level dependence within experimental groups and the contemporaneous
mitochondrial covariate make the reported p-values descriptive and add
application assumptions beyond intervention randomization.}

{The complete-data split has 13,475 training and 5,776 validation cells,
including 7,483 and 3,208 controls. Target-specific source arms contain
71--339 training and 30--146 validation cells. Every source cell belongs to a
GEM group represented among controls; depending on the target, 35--48 training
and 20--46 validation GEM groups contain both source and control cells. Across
the 378 eligible pairs, the ridge residual-variance ratio ranges from 0.860 to
0.991, so none is removed by the \(10^{-6}\) rule. These summaries establish
empirical overlap for the observed categorical groups but do not certify
positive conditional variance under arbitrary nonlinear adjustment. If the
support rule fails, PABM omits the assignment comparison for that pair.}

{A five-fold guide-stratified diagnostic fits the same held-out Poisson
perturbation--control association to every method-selected direction and its
reverse. Table~\ref{tab:k562-chr6-released} reports the resulting directional
statistics and within-fold BH summaries. Supplementary
Section~S1.4 gives the model, folds, standard
errors, and multiplicity rule.}
{Because the folds are cell-level and their training portions overlap,
this diagnostic measures within-panel stability rather than transport to a
new GEM group or experimental batch. A leave-group-out analysis would require
adequate source--control support for every retained intervention in each
training split and is not part of the reported study. The analysis also does
not isolate the effect of omitting mitochondrial percentage from adjustment.}

\begin{table}[!tbp]
\centering
\scriptsize
\begin{tabular}{lccccc}
\toprule
Procedure & \makecell{Constructed\\ edges} & \makecell{Outer-fold\\ edges} & \makecell{Median\\ \(|Z_{\mathrm{fwd}}|\)} & \makecell{Forward\\ stronger} & \makecell{Held-out\\ BH \(<0.05\)} \\
\midrule
{PABM} & 52 & 45.2 & 5.07 & 0.787 & 0.876 \\
ARGEN & 168 & 121.8 & 2.79 & 0.714 & 0.616 \\
\bottomrule
\end{tabular}
\caption{Chromosome-6 K562 comparison on 45 genes. Constructed
edges are obtained from the complete released panel. The outer-fold columns
average over five guide-stratified folds, each of which excludes its test
cells before graph selection. {PABM selects the {empirical residual-association
screen} and retains both comparisons in the complete fit and all five
outer-fold fits.} For each
selected direction, separate Poisson perturbation--control association
statistics are calculated in the displayed and reversed directions.
\(Z_{\mathrm{fwd}}\) is the forward association statistic, ``Forward
stronger'' is the fraction of selected pairs whose forward absolute statistic
exceeds its reverse absolute statistic, and the final column is the fraction with a held-out
Benjamini--Hochberg adjusted \(p\)-value below 0.05. These are post-selection
directional-association diagnostics conditional on the method-selected pairs.
Orientation accuracy and direct-edge recovery remain unidentified.}
\label{tab:k562-chr6-released}
\end{table}

{Supplementary Figure~S2 displays the
complete graphs. They share 28 arrows; two of 30 common adjacencies have
opposite orientations. PABM selects 52 complete-data arrows and 45.2 per outer
fold, versus 168 and 121.8 for ARGEN. ARGEN has greater pairwise fold overlap
{(mean pairwise Jaccard overlap 0.636 versus 0.588, where Jaccard overlap
is the number of shared directed edges divided by the number present in
either graph),} whereas PABM's selected pairs have larger
held-out directional-association summaries (Table~\ref{tab:k562-chr6-released}).
{A common query of STRING, a database of protein associations, supplies
undirected external evidence for} 13 of 52 PABM pairs and 31 of 168 ARGEN pairs,
versus 60 of all 990 panel pairs. These post-selection summaries address
sparsity, stability, and undirected functional compatibility; direct-edge and
orientation accuracy remain unidentified. Supplementary
Section~S1.4 gives the full protocol.}

{After removing PABM's 47 GEM indicators to align recorded covariates,
its graph remains smaller and its within-selected-pair association summaries
remain larger, while the STRING-supported fractions are nearly equal. This
sensitivity retains method-specific sample roles and selection and therefore
does not identify comparative causal accuracy.}

\section{Discussion}

{PABM separates two inferential tasks: valid interventions identify ancestry, while
proxy-adjusted response and intervention comparisons distinguish a direct parent
from an indirect ancestor. This separation permits target-specific conditional
distributions and different regression estimators within one graph-recovery
procedure. It also makes the identifying cost explicit. The model-based IV procedures compared here
impose models for measurements and edge coefficients; PABM imposes local
conditional-distribution restrictions for the candidate pairs. {Neither these strong, generally untestable requirements nor ARGEN's
assumptions generally imply the other. PABM's requirements allow} measurements informative about unmeasured confounding to enter
without a global parametric measurement model. Their plausibility must be
assessed for the regression covariates actually used.}

{Equation~\eqref{eq:imbalance-risk-gap} separates persistent nonparent
gain from score-estimation error. Better proxy adjustment addresses the first;
an appropriate working family, accurate fitting, and adequate validation size
address the second. The population framework allows distribution-sensitive
models, whereas the reported Gaussian and Poisson fits compare conditional
means. Detecting variance or shape effects requires corresponding estimators
and a distribution-sensitive ancestor procedure.}

{In the continuous simulations, omitting or permuting the proxy block
worsens SHD and directed F1 in every mechanism, and the reported PABM settings
have the best mean values of both metrics. Assignment evidence helps in two
designs and hurts in one. In the count study, PABM's automatic graphs have
higher directed F1 and much smaller SHD than ARGEN's, whereas ARGEN has higher
recall, average precision, and markedly better SHD at common graph sizes;
NOTEARS' near-empty selection is penalty-scale sensitive. These findings show
a selection--ranking tradeoff rather than uniform dominance.}

{Without a reference DAG, the K562 results remain descriptive: PABM gives
a smaller network and stronger held-out perturbation associations, ARGEN has
greater fold-to-fold overlap, and their STRING-supported fractions are similar
after input alignment. Independent experiments are needed to establish direct
effects and orientation. The Gaussian study likewise separates three issues:
exact adjustment permits exact conditional calibration, selected candidates
can induce positive nonparent gains, and noisy proxies can preserve residual
dependence even under exact ancestral selection.
{The proxy-removal comparisons evaluate the complete procedure, whereas
the Gaussian example isolates a failure of adjustment. Neither establishes
the adequacy of the recorded K562 covariates. Future studies should record
linked pre-intervention covariates or auxiliary measurements with a
scientifically justified adjustment role. Additional measurements or
separate assays do not by themselves guarantee the required independence;
their timing and possible responses to intervention must be considered.} Further work should develop
estimable sensitivity bounds, scientifically justified smaller regression
covariate sets, and estimator-specific calibration for flexible conditional
distributions.}

\section*{Acknowledgments and Disclosure of Funding}

This work was supported in part by NSF Grant DMS--2513668 and NIH
Grants R01AG069895, R01AG065636, and R01AG074858.
The authors declare no competing financial interests.

\bibliography{references}

\end{document}